\documentclass[11pt,a4paper,headinclude,footinclude,chapterprefix=on]{scrreprt}

\usepackage[utf8]{inputenc}
\usepackage[T1]{fontenc}
\usepackage{tabularx}
\usepackage{multicol}
\usepackage{graphicx}
\usepackage{acronym}
\usepackage{url}

\usepackage{tikz}
\usepackage{lipsum}

\usepackage[headsepline,plainheadsepline,footsepline,plainfootsepline]{scrpage2}

\usepackage{mathptmx}
\usepackage[scaled=.92]{helvet}
\usepackage{courier}

\usepackage[multiuser,nomargin,marginclue,footnote]{fixme}
\fxusetheme{color}
\fxsetup{targetlayout=colorcb}
\FXRegisterAuthor{awo}{anawo}{AWO}
\fxsetup{status=final}

\usepackage{geometry}
\usepackage[pscoord]{eso-pic}
\newcommand{\placetextbox}[3]{% \placetextbox{<horizontal pos>}{<vertical pos>}{<stuff>}
  \setbox0=\hbox{#3}% Put <stuff> in a box
  \AddToShipoutPictureFG{% Add <stuff> to current page foreground
    \put(\LenToUnit{#1\paperwidth},\LenToUnit{#2\paperheight}){\vtop{{\null}\makebox[0pt][c]{#3}}}%
  }%
}%

\usepackage{calc}
\usepackage{upgreek}
\usepackage{amsmath,amssymb,amsfonts,amsbsy,amsthm}
\usepackage{mathtools}
\usepackage{bbm}

\clearscrheadfoot

\ihead[\Large {\scshape TU Berlin }]{\Large {\scshape TU Berlin }}

\ifoot[{\tiny \begin{minipage}{4.0cm}Copyright at Technische Universität Berlin.\newline All Rights reserved.\end{minipage}}]{{\tiny \begin{minipage}{4.0cm}Copyright at Technische Universität Berlin.\newline All Rights reserved.\end{minipage}}}
\ofoot[Page \pagemark]{Page \pagemark}

\addtokomafont{pagefoot}{\normalfont}

\newcommand{\trnumber}{TKN-15-001}
\newcommand{\trdate}{February 2015}
\newcommand{\trauthor}{Konstantin Miller, Abdel-Karim Al-Tamimi,\\and Adam Wolisz}
\newcommand{\tremail}{\{miller,wolisz\}@tkn.tu-berlin.de, altamimi@yu.edu.jo}
\newcommand{\trtitle}{Low-Delay Adaptive Video Streaming Based on Short-Term TCP Throughput Prediction}

\begin{document}

\placetextbox{0.5}{0.99}{\fbox{\large{This TR is updated by TR TKN-16-001, available at \url{http://arxiv.org/abs/1603.00859}}}}%

%!TEX root = ../techreport.tex

% ================================================================
% Cover Sheet
% ================================================================
{
\sffamily

\thispagestyle{empty}

\setlength{\tabcolsep}{0pt}% remove indent within cell
\noindent% remove indent before cell
\begin{tabularx}{\columnwidth}{cXc}
  \includegraphics[height=1cm]{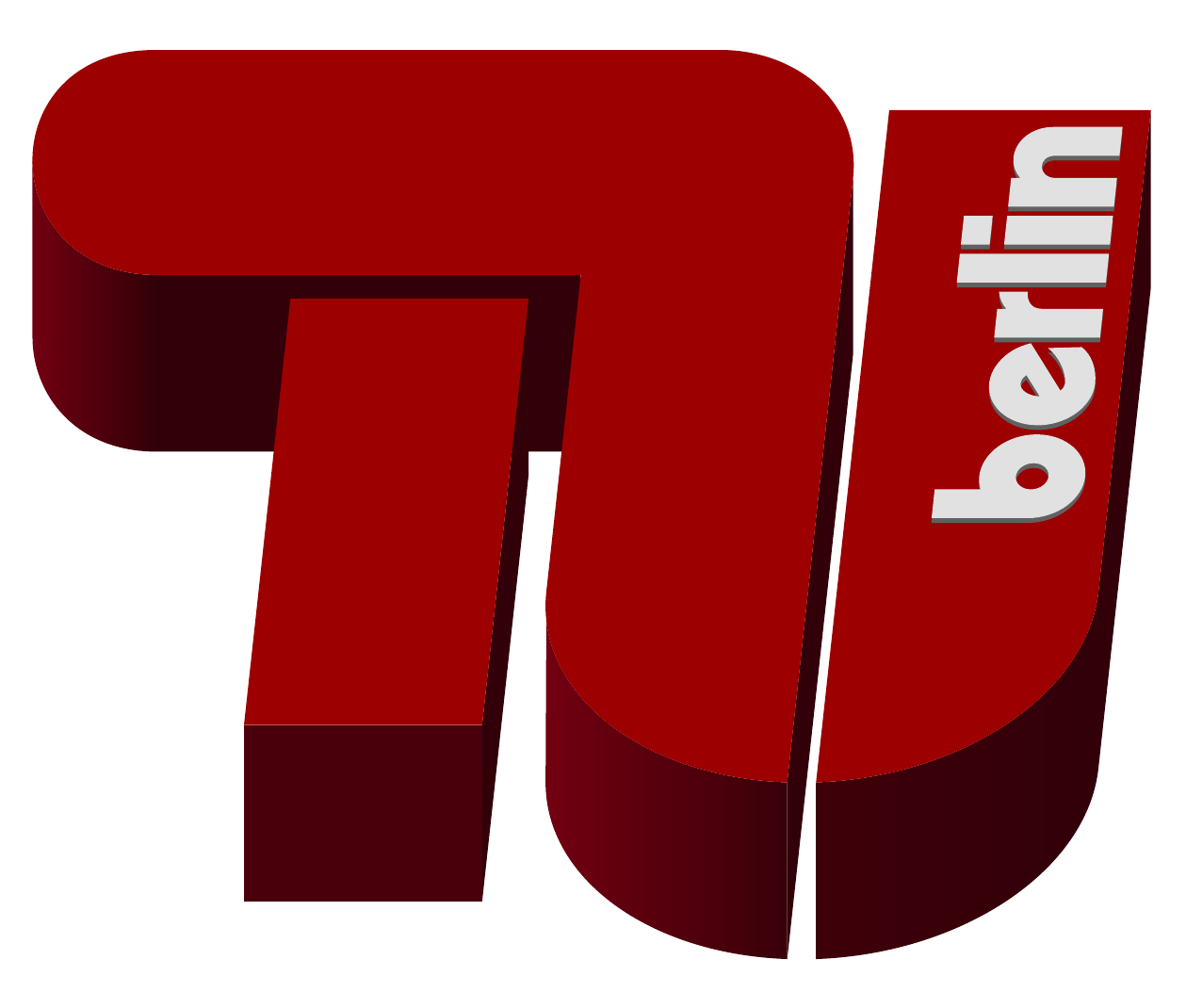}
  & &
  \includegraphics[height=1cm]{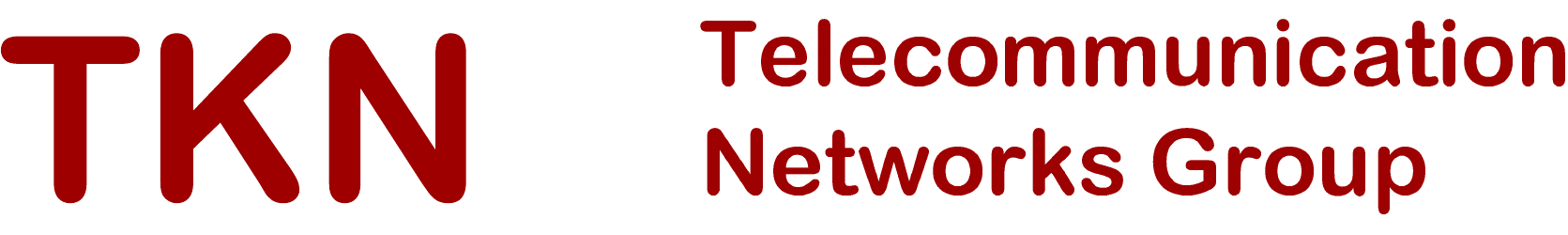}
  \\
\end{tabularx}
\setlength{\tabcolsep}{6pt}% back to default

\vspace{1.0cm}

\begin{center}
{\huge
\noindent
Technische Universität Berlin

\vspace{0.5cm}

\noindent
Telecommunication Networks Group

\begin{center}
\rule{15.5cm}{0.4pt}
\end{center}
}
\end{center}

\begin{minipage}[][11.0cm][c]{14.5cm}
{\Huge

\begin{center}
\trtitle
\end{center}

\begin{center}
{\LARGE \trauthor} \\
{\Large \tremail}
\end{center}

\begin{center}
Berlin, \trdate
\end{center}

\vspace{0.5cm}

}

\begin{center}
\setlength{\fboxrule}{2pt}\setlength{\fboxsep}{2mm}
\fbox{TKN Technical Report \trnumber}
\end{center}

\end{minipage}

\setlength{\fboxrule}{0.4pt}
\setlength{\fboxsep}{0.4pt}

\begin{center}

  \rule{15.5cm}{0.4pt}

  \vspace{0.5cm}

  {\huge {TKN Technical Reports Series}}

  \vspace{0.5cm}

  {\huge Editor: Prof. Dr.-Ing. Adam Wolisz}

  \vspace{0.5cm}

 \end{center}

}

\begin{abstract}
\section*{\abstractname}

Recently, \acl{HAS} has become the de facto standard for video streaming over the Internet. It allows the client to dynamically adapt media characteristics to varying network conditions in order to maximize \acl{QoE}, that is, minimize playback interruptions, while maximizing video quality at a reasonable level of quality changes. In the case of live streaming, where buffering possibilities are limited, this task becomes particularly challenging. An important factor than might help improving performance is the capability to correctly predict network throughput dynamics on short to medium timescales. This problem becomes notably difficult in wireless networks that are often subject to continuous throughput fluctuations.

In the present work, we develop an adaptation algorithm for \acl{HALS} that, for each adaptation decision, maximizes a \acl{QoE} based utility function depending on the probability of playback interruptions, average video quality, and the amount of video quality fluctuations. To compute the utility function, in particular the interruption probability, the algorithm leverages throughput predictions, and dynamically estimated prediction accuracy. 

We are trying to close the gap created by the lack of studies analyzing \acs{TCP} throughput on short to medium timescales. We study several time series prediction methods and model the distribution of prediction errors. We observe that \acl{SMA}, despite being the most straightforward method, performs best in most cases. We also observe that the relative underestimation error is best represented by a truncated normal distribution, while the relative overestimation error is best represented by a Lomax distribution. Moreover, although underestimations and overestimations are balanced in all traces, they exhibit a strong temporal correlation that we use to further improve prediction accuracy.

We compare the proposed algorithm with a baseline approach that uses a fixed margin between past throughput and selected media bit rate, and an oracle-based approach that has perfect knowledge of future throughput for a certain time horizon.

\end{abstract}

\tableofcontents

\chapter{Introduction}

Over the last years, we have been observing a dramatic change in video consumption patterns. The era of passive consumption of non-interactive "linear" content on a single device that does not offer much functionality is clearly over. Enabled by advances in media compression technologies, miniaturization and increasing processing power of electronic devices, complemented by ubiquitous availability of broadband access networks, we are witnessing the establishment of a new mindset: watch what I want, when I want, and where I want. A multitude of devices is at user's command to gain access to a vast sea of video content at any time and location: smartphones, tablets, PC's, game consoles, and, of course, TV sets that, however, underwent a transformation to become what is called Hybrid TV's, connected to the Internet and offering a multitude of interactive applications.~\cite{comScore2014a, Conviva2015} Moreover, as wearable devices such as smart watches and Google Glass begin to gain popularity, they might take the digital media landscape to a whole new level. All these devices empower the user to watch their favorite content on the best screen available at that moment, and not at the behest of the content provider.

Another game changer has been the advance of Web 2.0~\cite{O'Reilly2007}, that has expanded the focus of Internet from publishing to participation, allowing anyone to publish to the world without having to go through the closed systems that have dominated media since its very beginning. The plethora of freely accessible \ac{UGC}, shared over \aclp{SNS}, boosted the demand for watching video over the Internet and brought online video into the mainstream. As of today, the number of \ac{UGC} objects is orders of magnitude higher than that of traditional movies or TV programs and is rapidly evolving~\cite{Li2013a}. One prominent example reflecting this development is YouTube that alone reaches more US adults at the age of 18-34 than any cable network. As of January 2015, over 6 billion hours of video are watched each month on YouTube, which is almost an hour for every person on Earth. 100 hours of video are uploaded to YouTube every minute~\cite{YouTubeStats_medium}. And: around 25.0\% of the daily views on YouTube come from person-to-person sharing~\cite{Broxton2011}.

This development is being accompanied by a shift towards usage of wireless and mobile networks. In 2013, wired devices still accounted for the majority of Internet traffic at 56\%. The status quo, however, is rapidly changing. Traffic from wireless and mobile devices is predicted to exceed traffic from wired devices by 2018, accounting for 61\% of the total Internet traffic. And by far the largest part of it is video. Globally, video traffic is estimated to be 79\% of all consumer Internet traffic in 2018, up from 66\% in 2013~\cite{CiscoVNI2014}. The majority of streamed content is \ac{VoD}. However, the amount of live streaming, such as of sports events, video gaming, or music concerts, is experiencing rapid growth, promising significant revenues to the stakeholders~\cite{Gigaom2014}.

This enormous amount of traffic places a huge burden on the Internet infrastructure and on state-of-the-art wireless and mobile networks, and requires novel efficient solutions both in the area of wireless and mobile networking and video streaming. While a classical broadcaster exclusively uses a channel to broadcast to everyone within reach, on the Internet, the medium is shared among many users and a separate, unicast data stream is transmitted to every single receiver. Recent studies suggest that the challenges have not yet been successfully addressed. In 2013, around 26.9\% of streaming sessions on the Internet experienced playback interruption due to rebuffering, 43.3\% were impacted by low resolution, and 4.8\% failed to start altogether.~\cite{Conviva2014}

One of the problems is the Internet was not designed to support applications that require configurable end-to-end \ac{QoS}~\cite{ITU-T_E.800}. Considerable effort has been put into developing networking architectures, addressing this shortcoming~\cite{Aurrecoechea1998, Carapinha2010}. So far, none of them achieved a significant pervasiveness. Especially on wireless and mobile links, a user is exposed to interference, cross-traffic, and fading effects, leading to continuously fluctuating \ac{QoS} characteristics. As a consequence, we lately have been observing a thriving period for adaptive streaming technologies that are able to dynamically adjust the characteristics of the streamed media to varying network conditions, leading to a smoother viewing experience with less playback interruptions and a more efficient utilization of available network resources. 

In particular, one technology has become the de facto standard for Internet streaming: \ac{HAS}~\cite{Stockhammer2011a}. Its advantage is that the usage of \ac{HTTP} allows to leverage the ubiquitous and highly optimized \ac{HTTP} delivery infrastructure, including \acp{CDN}, caches, proxies, etc. This allows to reduce costs due to maintenance of specialized video servers. Also, \ac{HTTP} is usually allowed to traverse middleboxes, such as \ac{NAT} devices and firewalls. Of prime importance is also that \ac{HAS} has good scalability properties due to the stateless nature of \ac{HTTP} and because with \ac{HAS} the control logic resides within the client. Thus, the server is relieved from keeping extensive state, and maintaining persistent feedback loops with the client. An open standard, MPEG-DASH (Dynamic Adaptive Streaming over HTTP)~\cite{DASH2012, Sodagar2011}, has been introduced to facilitate interoperability.

An important feature of \ac{HAS} is that it uses \ac{TCP} to transport the data, which has its pros and cons. On the one hand, \ac{TCP} offers built-in congestion control and congestion avoidance mechanisms, that are necessary to maintain network stability, as well as to ensure basic fairness among competing flows. It also offers reliable communication by retransmitting lost packets, which enables usage of efficient video compression technologies that are particularly sensitive to packet losses. These features allow to reduce the complexity of the streaming application, which otherwise would have to implement mechanisms to deal with losses and congestion.~\cite{Zhang2001} On the other hand, retransmissions delay subsequent packets, making \ac{TCP} more challenging for live streaming. Further, \ac{TCP} reacts to packet losses and transmission delay peaks by reducing its sending rate, which might negatively impact video quality.

The operation of \ac{HAS} can be roughly described as follows. The video material is encoded in several representations, that vary w.r.t. their media characteristics such as spatial resolution, frame rate, compression level, etc. Typically, different representations have different media bit rates and thus different requirements on network throughput. Representations are split into segments, typically containing one to few seconds of video data. The data is encoded such that segments from different representations can be seamlessly played back after each other. The client issues a series of \ac{HTTP} requests to download the segments in appropriate representations, trying to achieve certain goals. 

For video streaming, identifying performance goals and expressing them in a way that facilitates objective measurement has turned out to be an extremely challenging task. One of the main reasons is that the ultimate target of a streaming service is a human being. Thus, an evaluation of such a service must inevitably take into account human perception and cognitive processing, involved in consuming the video content. These phenomena, however, are influenced by a large amount of hardly measurable factors. The notion of \acf{QoE} was introduced in an effort to assess these phenomena and help making them accessible to an objective evaluation process. The \ac{ITU} defines \ac{QoE} as "the overall acceptability of an application or service, as perceived subjectively by the end-user", which might be influenced by "user expectations" and "context"~\cite{ITUT_P10G100_Am2}. The number of factors influencing \ac{QoE} is immense, and many of them have a high level of subjectivity, which results in extremely complex modeling.~\cite{Reiter2014}
%(See Chapter~\ref{cha:qoe} for more details.)

%, such as to optimize user's \ac{QoE}~\cite{Juluri2015}, achieve a certain level of fairness in multi-user environments, avoid wasting resources and generating unnecessary costs, etc.

With \ac{HAS}, the degrees of freedom for maximizing \ac{QoE} are determined by the choice of \ac{TCP} as transport protocol on the one hand, and by putting the adaptation logic into the client, on the other. The media characteristics of available video representations are configured by the service provider during the planning phase.~\cite{Seufert2014} The main factors influencing \ac{QoE} that can be controlled by a \ac{HAS} client are: initial delay, number and duration of rebuffering periods, selected video representation, and number and amplitude of representation changes. Their relative importance for \ac{QoE} is, however, still poorly understood. Nevertheless, the number of studies dedicated to this topic has been dramatically increasing with the growing importance of video streaming, so that a lot of valuable insights are available to help designing \ac{QoE}-optimized streaming mechanisms. 

In particular, many studies suggest that the number and duration of rebuffering periods have the most severe impact on \ac{QoE}, especially with live streaming.~\cite{Conviva2014} In particular, users are willing to accept a higher initial delay and higher video distortion due to increased compression rate, if it helps minimizing rebuffering periods.~\cite{Seufert2014, Hossfeld2012a, Quan2008, Singh2012a} On the other hand, it was observed that video quality fluctuations resulting from dynamically changing the representation can have a negative impact on \ac{QoE}.~\cite{Lewcio2011, Yitong2013} In particular, some studies come to the conclusion that a lower overall video quality might be tolerated if it helps reducing the amount of representation changes.~\cite{Pessemier2013}

In the present study, we take up the position that a crucial factor influencing the ability of the client to maximize \ac{QoE}, in particular, to minimize rebuffering, is his capability to correctly estimate network throughput dynamics on short to medium timescales. Specifically in the case of live streaming, where the time horizon for prefetching segments is limited and the time between the moment when a segment becomes available for download and its playback deadline typically constitutes few seconds, a client can strongly benefit from having a precise estimation of network throughput. This task is particularly challenging in wireless and mobile networks. It is further complicated by \ac{TCP}'s congestion avoidance and control feedback loop, as well as retransmission mechanisms, contributing to the complexity of application-layer throughput dynamics.

\textbf{Our contribution.} In our work, we turn out attention to designing an adaptation algorithm for \ac{HALS}. Our idea is, prior to each segment download, to compare potential future adaptation trajectories and to select the one maximizing \ac{QoE}. In order to evaluate \ac{QoE} of an adaptation trajectory, we define a utility function depending on three factors that we call subutilities: (i) probability that a segment misses its playback deadline and thus the streaming session enters a rebuffering period, (ii) the distortion of video, evaluated by means of \ac{PSNR}, and (iii) the number and amplitude of representation changes. The utility is computed from individual subutilities in accordance with available literature on \ac{QoE}. In particular, rebuffering subutility appears as a multiplicative factor and plays the role of an upper bound on the total utility, reflecting its strong impact on \ac{QoE}. The other factor is a weighted sum of subutilities representing distortion and quality fluctuations that allows to resolve their trade-off in a configurable way.

In order to compute the defined utility, in particular, the probability that a segment misses its playback deadline, we study the predictability of \ac{TCP} throughput in wireless networks over timescales from 1 to 10 seconds. For our study we use \ac{TCP} throughput traces collected in IEEE 802.11bg \acp{WLAN} throughout Berlin, Germany, including public hotspots (indoor and outdoor), campus hotspots, and access points in residential environments. In particular, we focus on traces with low average throughput (hundreds of kilobits to few megabits per second), and high throughput fluctuations, which make the operation of a streaming client particularly challenging.
We evaluate different time series prediction methods using varying numbers of past throughput measurements. We demonstrate that 
the most na\"{\i}ve method, \ac{SMA} outperforms more sophisticated methods on all timescales, independent of the specific throughput dynamics.
This means, somewhat surprisingly, that in studied environments, accounting for the trend in the past measurements does not help to increase prediction accuracy.

We further observe that prediction accuracy strongly varies across studied traces. Therefore, it is inefficient to assume a fixed prediction error and account for this error by a fixed margin between predicted throughput and selected media bit rate. 
On the contrary, it is crucial to dynamically estimate the prediction accuracy, in order to allow clients to efficiently utilize available network resources, at the same time being robust to throughput fluctuations. 
Consequently, we study approaches to model the prediction error and to estimate it for individual streaming sessions. 
We demonstrate that the overestimation error is extremely well represented by the Lomax distribution~\cite{Johnson1994} on all considered timescales. The underestimation error is best represented by a truncated normal distribution except for the timescale of 1 second, where the truncated logistic distribution results in a slightly better Kolmogorov-Smirnov distance~\cite{Hollander2014} between the empirical and the theoretical \ac{CDF}. In addition, we find out that although underestimations and overestimations are balanced over the total duration of individual traces, they exhibit a strong temporal correlation that can be used to further improve prediction accuracy.

Armed with these insights we develop a novel adaptation algorithm for live streaming, which takes into account throughput predictions and an estimation of the relative prediction error, in order to maximize the defined \ac{QoE}-related utility function. We evaluate the developed algorithm using collected throughput traces and show that it outperforms the baseline approach which uses a fixed margin.

In the following, 
Chapter~\ref{sec:related_work} presents related work, 
Chapter~\ref{sec:system_model} describes our system model, 
Chapter~\ref{sec:traces} details collected throughput traces, 
Chapter~\ref{sec:prediction} studies throughput predictability and prediction accuracy,
Chapter~\ref{sec:adaptation} presents the developed adaptation algorithm,
Chapter~\ref{sec:evaluation} evaluates its performance,
and Chapter~\ref{sec:conclusion} concludes the paper.

\chapter{Related work}
\label{sec:related_work}

% \cite{Lohmar2011, Stockhammer2012, Concolato2013, Wei2014}
%HTTP Adaptive Streaming in practice, Mark Watson, Proceedings of the ACM MMSys Conference, 2011
%Adaptive Bit Rate video delivery, Thomas Kernen, Cisco, 2013 , DASH265 / HEVC Page 19 

There is a lack of studies systematically investigating approaches for \ac{TCP} throughput prediction on short to medium timescales and evaluating their prediction accuracy. Consequently, only few works present adaptation algorithms that explicitly take into account throughput predictions. A common approach followed by many adaptive streaming clients is to use throughput averaged over a number of past measurements and one or multiple fixed safety margins to make adaptation decisions~\cite{MillerK2012, Jiang2012, Liu2011}. Other approaches leverage bandwidth probing techniques to obtain an estimation of network throughput~\cite{Mok2012}, which, however, require support from the network infrastructure, server instrumentation, and/or modifying lower protocol layers.

Similar in spirit to our work is the rationale behind the adaptation algorithm proposed by Liu and Lee~\cite{Liu2014a}. In contrast to their work, however, we first study statistical properties of prediction errors, which allows us to design an adaptation algorithm that uses a parametric approach, fitting the \ac{CDF} of the prediction error to a distribution type determined during the preceding study. Proceeding this way requires significantly less data which allows the algorithm to operate without a database of measurements collected in the same environment. In addition, we separately model prediction errors on different timescales from 1 to 10 seconds. Moreover, we separately model underestimation and overestimation errors, which have quite different distributions, and their temporal correlation, which allows us to further improve prediction accuracy. Finally, instead of enforcing a minimum time between video quality adaptations, we are maximizing a utility function that includes a quality fluctuations related term, which is a more flexible approach, better suitable for live streaming, and allowing for a higher resource utilization.

Yin et al.~\cite{Yin2014} study the effect of prediction errors on performance of three adaptation approaches, buffer based, rate based, and \ac{MPC}. The study is using synthetic throughput traces and, due to the lack of literature on the subject, assumes that the prediction error has a normal distribution. In our study, we try to close this gap by evaluating several prediction methods and modeling the achievable prediction error on different timescales. 
%We demonstrate that the assumption of a normally distributed prediction error is not always justified. 
Further, we develop an adaptation algorithm which takes into account the buffer level, throughput predictions on several timescales, and dynamically estimated prediction accuracy, and which specifically targets live streaming.

Tian and Liu~\cite{Tian2012} propose a prediction-based adaptation algorithm, where the media bit rate is selected to equal predicted throughput times dynamically varying adjustment factor. In contrast to this work, we predict throughput separately on different timescales, dynamically estimate prediction accuracy, and leverage temporal correlation of underestimations and overestimations. In addition, we select a video representation by maximizing a \ac{QoE}-based utility function, depending on buffer underrun probability, video quality and video quality fluctuations.

Wang et al.~\cite{Wang2008} argue that for best performance, the mean media bit rate of streamed video should be roughly half the available network throughput. In our work we study throughput prediction accuracy in different environments, which enables us to develop an algorithm that dynamically tunes the margin between selected media bit rate and predicted throughput based on the buffer level and estimated prediction accuracy.

Jarnikov et al.~\cite{Jarnikov2011a} propose an adaptation algorithm based on a Markov decision process. With this approach, an optimal strategy is calculated offline for a given throughput distribution. We argue that assuming a fixed distribution, and neglecting temporal correlations, does not properly account for the variability of throughput dynamics in different environments.

Further improvement of an algorithm such as the one proposed in this work can potentially be achieved by complementing it with a data-driven, potentially location-based, approach operating as an outer loop on a macroscopic level~\cite{Liu2012a, Riiser2012, Hao2014}.

A important research topic is \ac{QoE} for adaptive video streaming~\cite{Balachandran2012, Seufert2014, Reiter2014, Song2014}. Rebuffering, initial delay, and quality fluctuations are factors that have not been part of traditional \ac{QoE} metrics for video, but that have a tremendous impact on user's perception of adaptive video streamed over a best-effort network, such as the Internet. User engagement is another important metric, which is especially of interest for content providers since it is directly related to advertising-based revenue schemes~\cite{Balachandran2013, Conviva2014}.

\chapter{System model and notation}
\label{sec:system_model}

\begin{figure}[t]
\centering
\includegraphics[scale=1.1]{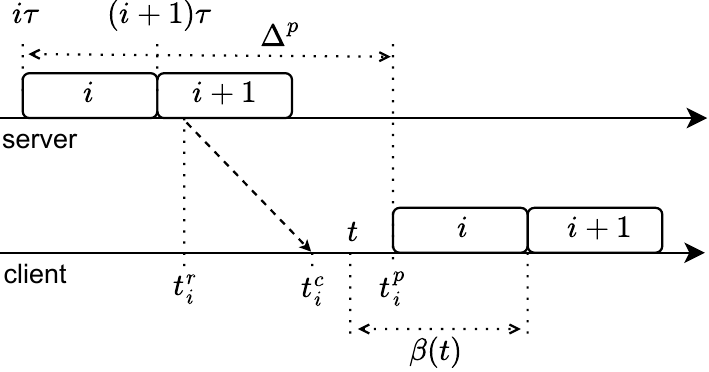}
\caption{Illustration of client and server (event) timelines.}
\label{fig:system_model}
\end{figure}

%Roughly speaking, the operation of \ac{HAS} clients such as those following the open standard MPEG-DASH~\cite{DASH_specification}, or proprietary solutions such as Microsoft Smooth Streaming, \ac{HLS}, or Adobe Dynamic Streaming, can be described as follows (see, e.g.,~\cite{Stockhammer2011a} for more details). With \ac{HAS}~\cite{Stockhammer2011a}, the video material is encoded in several representations, that may vary w.r.t. their spatial resolution, frame rate, compression level, etc. Typically, different representations have different media bit rates and thus different requirements on network throughput. The individual representations are split into segments, typically containing one to few seconds of video data. The data is encoded such that segments from different representations can be seamlessly played back after each other. Prior to starting the streaming session, a video client downloads a \ac{MPD} file containing information about the individual segments and their \acp{URL}. The client issues a series of \ac{HTTP} requests to download the segments in appropriate representations, trying to achieve certain goals, such as to optimize user's \ac{QoE}, achieve a certain level of fairness in multi-user environments, avoid wasting resources and generating unnecessary costs, etc.

In the considered scenario a \ac{HAS} client is live-streaming an event. The event to be streamed shall start at $t=0$. The stream is partitioned into segments. We denote the duration of video content contained in one segment by $\tau$. In order to simplify the presentation, we assume that all segments have equal duration. We use index $i\in\{1,2,\ldots,n\}$ to indicate a particular segment in a stream.  % All presented solutions can, however, be easily extended to the general case.

Segment $i$ shall contain video material covering the time period $\left[i\tau,\,(i+1)\tau\right]$, and become available for download at time $(i+1)\tau$. We assume that a segment must be completely downloaded prior to being processed by the client, and that the processing time at the client (that is, demultiplexing and decoding) is equal to 0, that is, the playback of a segment can start immediately after it has been completely downloaded. We argue that these fixed delays can be omitted without loss of generality. Using fixed non-zero delays would not affect the results but would make the notation more cumbersome.

Each segment is available in several representations. We denote the set of available representations by $\mathcal{R}$, indexed by $j\in\left\{1,2,\ldots,m\right\}$, with $m=\lvert\mathcal{R}\rvert$. W.l.o.g., $\mathcal{R}$ shall contain only representations feasible for the considered user. (A representation might be infeasible if its playback requires features not supported by the user or if its properties are excluded by configuration.)

We denote by $s_{ij}$ the size in bits, and by $\bar{r}_{ij}=s_{ij}/\tau$ the \ac{MMBR} of segment $i$ from representation $j$. 
We denote by $\bar{r}_j=1/n\sum_{i=1}^n{\bar{r}_{ij}}$ the \ac{MMBR} of representation $j$.
%We write $\bar{r}_{\min}=\min_{j=1}^{m}{\bar{r}_j}$ ($\bar{r}_{\max}=\max_{j=1}^{m}{\bar{r}_j}$) for the lowest (highest) \ac{MMBR} of a representation in $\mathcal{R}$.
If the representation of a segment is clear from the context, we might omit index $j$. Thus, $s_i$ might, e.g., denote the size of a downloaded segment $i$,  from the representation that was used to download it. Consequently, $\bar{r}_i=s_i/\tau$ then denotes the \ac{MMBR} of segment $i$.

We use the following real-valued variables to denote continuous time in seconds. $t^r_i$ denotes the time when the request to download segment $i$ is sent by the user (at $t_i^r$, the client either just finished downloading the previous segment $i-1$, or segment $i$ just became available at the server). $t_i^c$ denotes the time when the last bit of segment $i$ is received by the user. $t_i^p$ denotes the time when the playback of segment $i$ is started. See Figure~\ref{fig:system_model} for an illustration.

We denote the maximum playback delay by $\Delta^p_{\max}\geq 2\tau$, that is, playback of segment $i$ must not start later than $i\tau+\Delta^p_{\max}$. The lower bound of $2\tau$ stems from the fact that a segment can only be published $\tau$ seconds after its start, and that it takes, on average, up to further $\tau$ seconds to download it. Typical values for $\Delta^p_{\max}$ are assumed to be around 2 to 10 seconds. The actual playback delay, $\Delta^p\in\left[2\tau,\,\Delta^p_{\max}\right]$, is determined by the initial delay during the start of the streaming session and can be readjusted during each rebuffering event. The start of playback of segment $i$ is thus given by $t_i^p=i\tau+\Delta^p$. The value of $\Delta^p$ determines the maximum attainable buffer level, given by $\Delta^p-\tau$, and thus determines client's sensitivity to throughput fluctuations and potential link outages. A client might dynamically tune $\Delta^p$ based on estimated throughput and link outage statistics. We leave that for future work and assume $\Delta^p=\Delta^p_{\max}$, maximizing the attainable buffer level, defined by 
\begin{equation*}
\beta(t)=\max\left\{t_i^p\;|\;t_i^c\leq t\right\}+\tau-t\,,
\end{equation*}
which is the time until all segments downloaded until time $t$ are played back, see Figure~\ref{fig:system_model}.

We denote by $\rho\left(t_1,t_2\right)$ the mean application layer throughput in the time interval $\left[t_1,t_2\right]$, that is, the number of bits received by the application from the \ac{TCP} layer during this time interval, divided by $t_2-t_1$.

%We assume that every second, the clients computes an estimation of future throughput for periods of duration $T\in\{1,\ldots,T_{\max}\}$ with $T_{\max}=10$ s. We denote by $\hat{\rho}\left(t_1,\,t_2\right)$ the most recent throughput prediction for the time interval $\left[t_1,\,t_2\right]$.

% If at time $t$ buffer level is $0\leq\beta(t)\leq B_{\max}$, information about future segment sizes is available for $B_{\max}-\beta(t)$ s. Use prediction for segment sizes from Karim's work?

\chapter{\acs{TCP} throughput traces}
\label{sec:traces}

\begin{figure}
\centering
\includegraphics[scale=1]{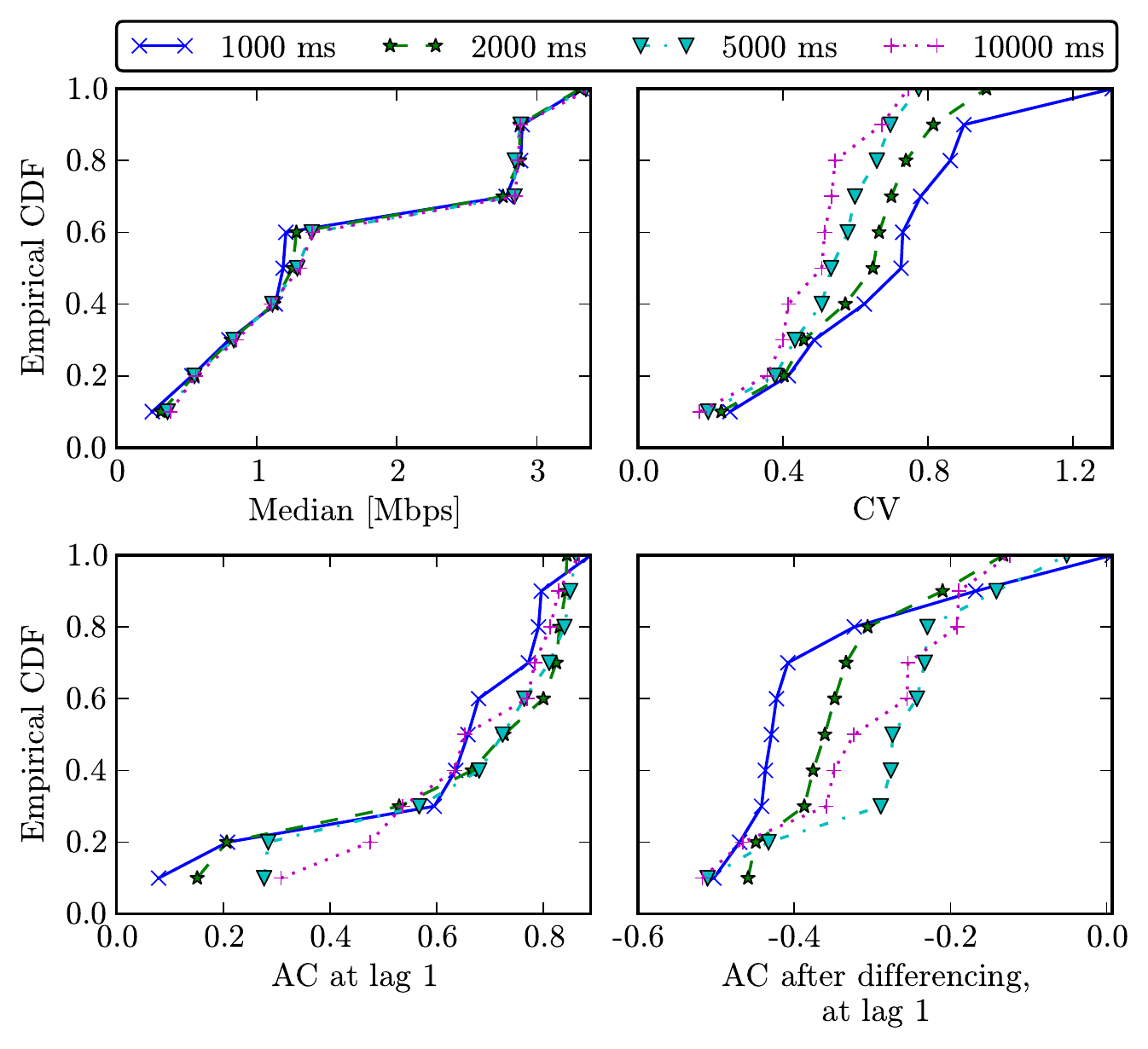}
\caption{Application layer throughput statistics: median of the throughput, \ac{CV}, auto-correlation at lag 1, and auto-correlation, after differencing, at lag 1. Lag 1 means here that the lag equals the size of the sampling interval.}
\label{fig:trace_statistics}
\end{figure}

For our study of \ac{TCP} throughput prediction, as well as for performance evaluation of the developed adaptation algorithm, we recorded \ac{TCP} throughput traces in IEEE 802.11bg networks at different locations throughout Berlin, Germany, including public hotspots (indoor and outdoor), campus hotspots, and access points in residential environments. The duration of the traces varies between 1500 and 1800 seconds each. The traces were collected using Lenovo ThinkPad L430 and T420 laptops, running Ubuntu 13.04 and Ubuntu 14.04 operating systems, with default \ac{MAC} and \ac{TCP} configurations.

From the set of collected traces we selected 10 that have a mean application-layer throughput between few hundreds of kilobit and few megabit per second. This throughput does not allow the client to select highest video representation, which we assume to be \ac{HD} video with an \ac{MMBR} around 4 to 7 Mbps~\cite{Krishnappa2013, Gigaom2014}. In addition, the selected traces exhibit high throughput fluctuations that make low-delay live streaming particularly challenging.

From the traces, we generated time series containing incoming packets statistics, computed over sliding time intervals of 1 s to 10 s duration, shifted with a step size of 1 s. In addition to throughput statistics, our traces contain internal \ac{TCP} and internal \ac{MAC} information. 
The latter have been collected by putting the network interface into monitor mode and using radiotap headers~\cite{radiotap}. \ac{TCP} information includes delay jitter statistics and statistics of outstanding bytes. \ac{MAC} information includes number of own frames received, number of other frames received, modulation scheme statistics, \ac{SSI} statistics, and retransmission statistics.

To give a rough idea about properties of individual traces, Figure~\ref{fig:trace_statistics} presents basic throughput statistics. It shows median throughput, \ac{CV}, auto-correlation at lag 1, and auto-correlation after differencing, at lag 1. Lag 1 means here that the lag equals the size of the averaging interval. The values are presented as \acp{ECDF}, where each point in the graph corresponds to one trace. 
All traces can be downloaded from \url{http://ns.tkn.tu-berlin.de/miller/}. The software used to collect and process the data can be handed out upon request.

% TODO: (Stationarity. How do those statistics vary over time.)
% TODO: (Spectrum density?)

\chapter{Short-Term TCP throughput prediction}
\label{sec:prediction}

In this chapter, we present results on \ac{TCP} throughput prediction for timescales from 1 to 10 seconds. Section~\ref{sec:methods} describes a selection of studied time series prediction methods, Section~\ref{sec:performance} presents evaluation results for three simple methods: \ac{SMA}, linear extrapolation, and double exponential smoothing (Holt-Winters), and Section~\ref{sec:modeling} presents modeling of the relative prediction error.

\section{Prediction methods}\label{sec:methods}

%\begin{figure*}
%\centering
%\includegraphics[scale=0.95]{fig/err_rel_cdf_all_traces_as_one}
%\caption{Comparison of prediction errors for different prediction methods on different timescales. (Higher values indicate higher prediction accuracy.) See Section~\ref{sec:performance} for details.}
%\label{fig:err_rel_cdf_all_traces_as_one}
%\end{figure*}

%\begin{figure*}
%\centering
%\includegraphics[scale=1]{fig/compare_quantiles}
%\caption{Distribution of relative prediction error quantiles for different predictions methods over individial traces. See Section~\ref{sec:performance} for details.}
%\label{fig:compare_quantiles}
%\end{figure*}

We evaluated a number of time series prediction techniques, including \ac{SMA}, linear extrapolation, \ac{CSS}, several flavors of exponential and double exponential smoothing, \ac{ARIMA}, machine learning based methods, etc. A selection of studied methods is briefly described in the following. We abbreviate the methods by $\langle$type$\rangle$:$\langle n\rangle$:$\langle$parameters$\rangle$, where $\langle$type$\rangle$ is the name of the method, $n$ is the number of past measurements used as input, and $\langle$parameters$\rangle$ include further optional configuration parameters.

% neither the increasing the number of past samples taken into account nor

\subsection{Simple moving average}

\ac{SMA} is probably the most simple prediction method imaginable. The predicted value is the average over a number of past measurements. The configuration parameters are: the number of past measurements, and the type of used mean value: arithmetic, geometric, or harmonic. In the following, we abbreviate this method with SMA:$\langle n\rangle$:$\langle$mean type$\rangle$, where $n\geq 1$ is the number of past measurements and type is one of $\{\text{ar},\text{gm},\text{hm}\}$. SMA:2:ar, e.g., means that the predicted value is the arithmetic mean from two past measurements. In particular, we denote the na\"{\i}ve approach of using the most recent measurement as predicted value with SMA:1:ar.

\subsection{Simple exponential smoothing}

With \ac{SES}, the predicted value is computed by averaging the past measurements, exponentially decreasing weights of older measurements. For given past measurements $x_1,\ldots,x_n$, the predicted value is computed as $x_{n+1}=a_n$, with
\begin{align*}
a_1&=x_1\\
a_k&=\alpha x_k + (1-\alpha)a_{n-1}\,.
\end{align*}
Besides the number of past measurements, it has a configuration parameter $\alpha\in\left[0,1\right]$. We tune $\alpha$ for each prediction by minimizing the \ac{MSE} within past measurements, given by
\begin{equation*}
\frac{1}{n-1}\sum_{k=2}^{n}{\left(x_k-a_{k-1}\right)^2}\,.
\end{equation*}
We abbreviate \ac{SES} with \ac{SES}:$\langle n\rangle$:mse, where $n\geq 2$ is the number of past measurements, and "mse" indicates the approach used to tune $\alpha$.

\subsection{Linear extrapolation}

Linear extrapolation is another straightforward prediction method that differs from \ac{SMA} in that it takes into account the linear trend from past measurements. More specifically, linear extrapolation fits a linear curve into the set of given past measurements, minimizing the mean square error, and computes the prediction from extrapolating the curve to the prediction horizon. It thus requires at least two past measurements to compute a prediction. We abbreviate linear extrapolation with LinExt:$\langle n\rangle$, where $n\geq 2$ is the number of past measurements.

\subsection{Double exponential smoothing}

Similar to linear extrapolation, double exponential smoothing tries to account for the trend in the data, however, it assumes that most recent measurements have a higher significance for the prediction, and assigns older measurements exponentially decreasing weights. In the following, we use a variant of the method, sometimes called Holt-Winters double exponential smoothing. With Holt-Winters, for given past measurements $x_1,\ldots,x_n$, the prediction is computed as $a_n + b_n$, where $a_n,\;b_n$ are computed by the following recursive procedure.
\begin{align*}
a_2&=x_2\\
b_2&=x_2-x_1\\
a_k&=\alpha x_k + (1-\alpha)(a_{k-1}+b_{k-1})\,,\;\text{for}\;k>2\\
b_k&=\beta(a_k-a_{k-1})+(1-\beta)b_{k-1}\,,\;\text{for}\;k>2\,.
\end{align*}

The Holt-Winters method has configuration parameters $\alpha$ and $\beta$ that dramatically influence the prediction quality and thus have to be carefully tuned. In our work we tune them for each prediction by minimizing the \ac{MSE} within past measurements, which is given by
\begin{equation*}
\frac{1}{n-2}\sum_{k=3}^{n}{\left(x_k-(a_{k-1}+b_{k-1})\right)^2}\,.
\end{equation*}
Thus, this method requires at least three past values for the prediction. As abbreviation we use HW:$\langle n\rangle$:mse, where $n\geq 3$ is the number of last values, and "mse" indicates the approach used to tune $\alpha$ and $\beta$.

\subsection{Cubic smoothing splines}

\ac{CSS} model provides both smooth historical trend and a linear prediction function. The method uses a likelihood approach to estimate the smoothing parameter. It is based on finding piecewise cubic polynomials that are joined at the equally spaced time series points.~\cite{Hyndman2005a}

\subsection{\acf{LOESS}}

\ac{LOESS} is a non-parametric regression method that tries to build a model describing the deterministic part of the variation in the data by incrementally fitting localized subsets of the data using simple regression models.

\subsection{Autoregressive model}

Autoregression model predicts the variable of interest by using a linear combination of past measurements, plus white noise. For past measurements $x_1,\ldots,x_n$, the prediction of an autoregressive model of order $p$ can be written as:
\begin{equation*}
x_{n+1} = c + \sum_{i=1}^n \alpha_i x_i + \omega_n\,,
\end{equation*}
where $c$ is a constant and $\omega_n$ denotes white noise. The order $p$ is selected by optimizing the \ac{AIC}.  

\subsection{\acf{ARIMA}}

\ac{ARIMA} is a combination of autoregressive and moving average models, with the ability to use a differenced or integrated representation of the time series. In our study, we were using a statistical method proposed in \cite{Hyndman2007} that uses a combination of unit root test, minimization of the \ac{AIC} and \ac{MLE} to reach an optimized \ac{ARIMA} model.

\section{Evaluation of prediction accuracy}\label{sec:performance}

\begin{figure}
\centering
\includegraphics{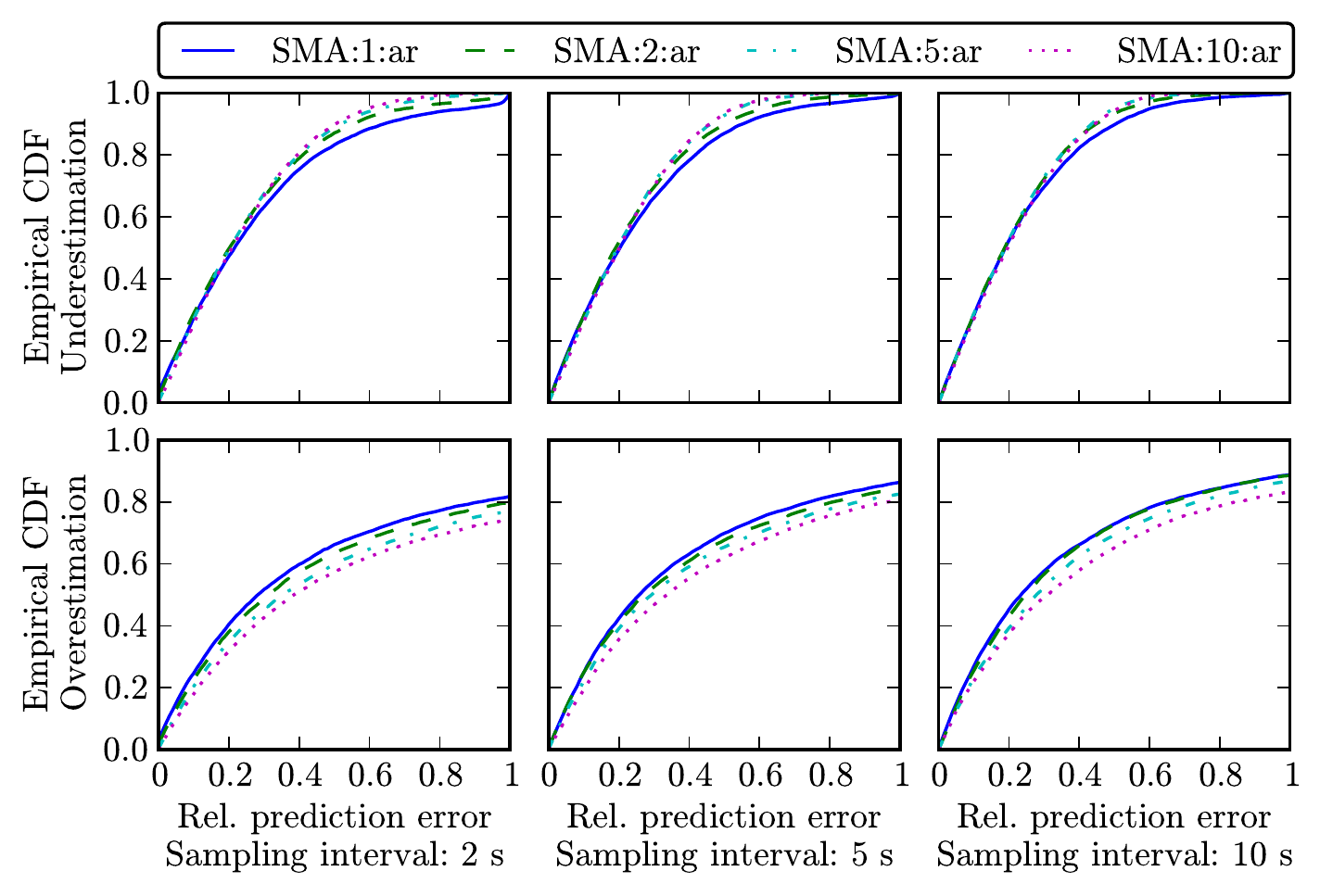}
\caption{Relative prediction error for \ac{SMA} on different timescales. (Higher values indicate higher prediction accuracy.) See Section~\ref{sec:performance} for details.}
\label{fig:err_rel_cdf_all_traces_as_one_00}
\end{figure}

\begin{figure}
\centering
\includegraphics{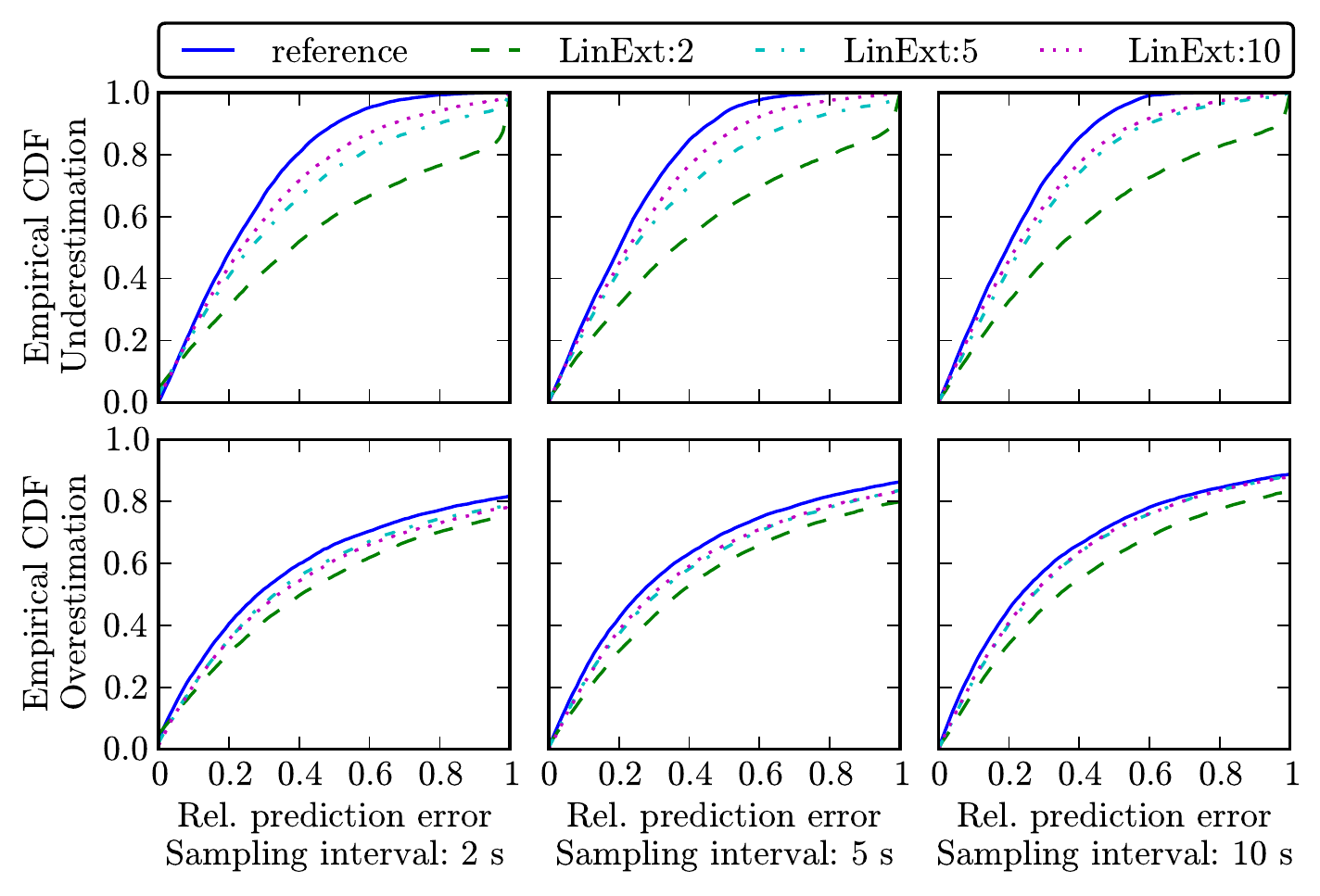}
\caption{Relative prediction error for linear extrapolation on different timescales. (Higher values indicate higher prediction accuracy.) See Section~\ref{sec:performance} for details.}
\label{fig:err_rel_cdf_all_traces_as_one_01}
\end{figure}

\begin{figure}
\centering
\includegraphics{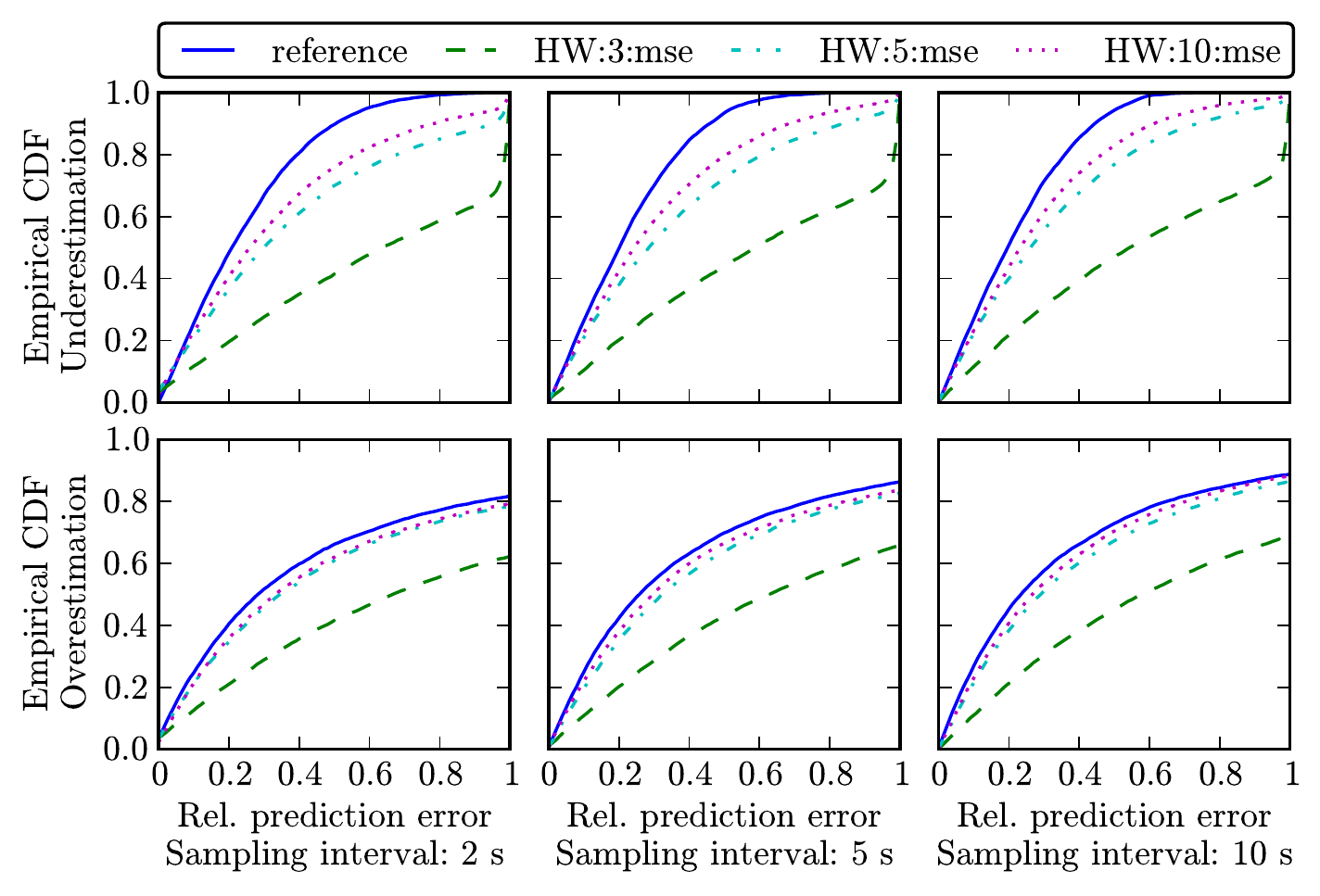}
\caption{Relative prediction error for double exponential smoothing (Holt-Winters) on different timescales. (Higher values indicate higher prediction accuracy.) See Section~\ref{sec:performance} for details.}
\label{fig:err_rel_cdf_all_traces_as_one_02}
\end{figure}

\begin{figure}
\centering
\includegraphics{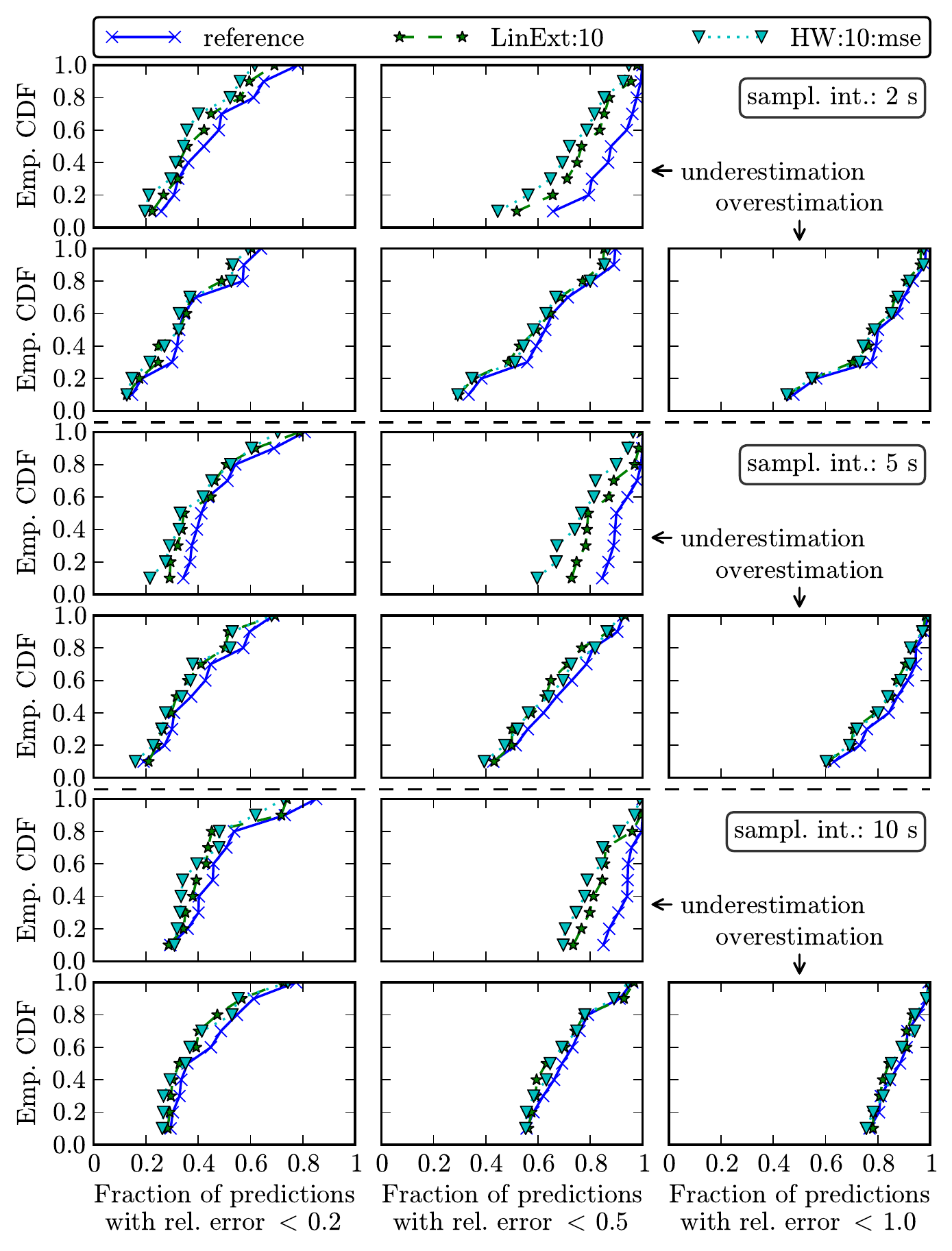}
\caption{Quantiles of the relative prediction error for different prediction methods in individual traces. See Section~\ref{sec:performance} for details.}
\label{fig:compare_quantiles}
\end{figure}

\begin{figure}
\centering
\includegraphics[scale=1]{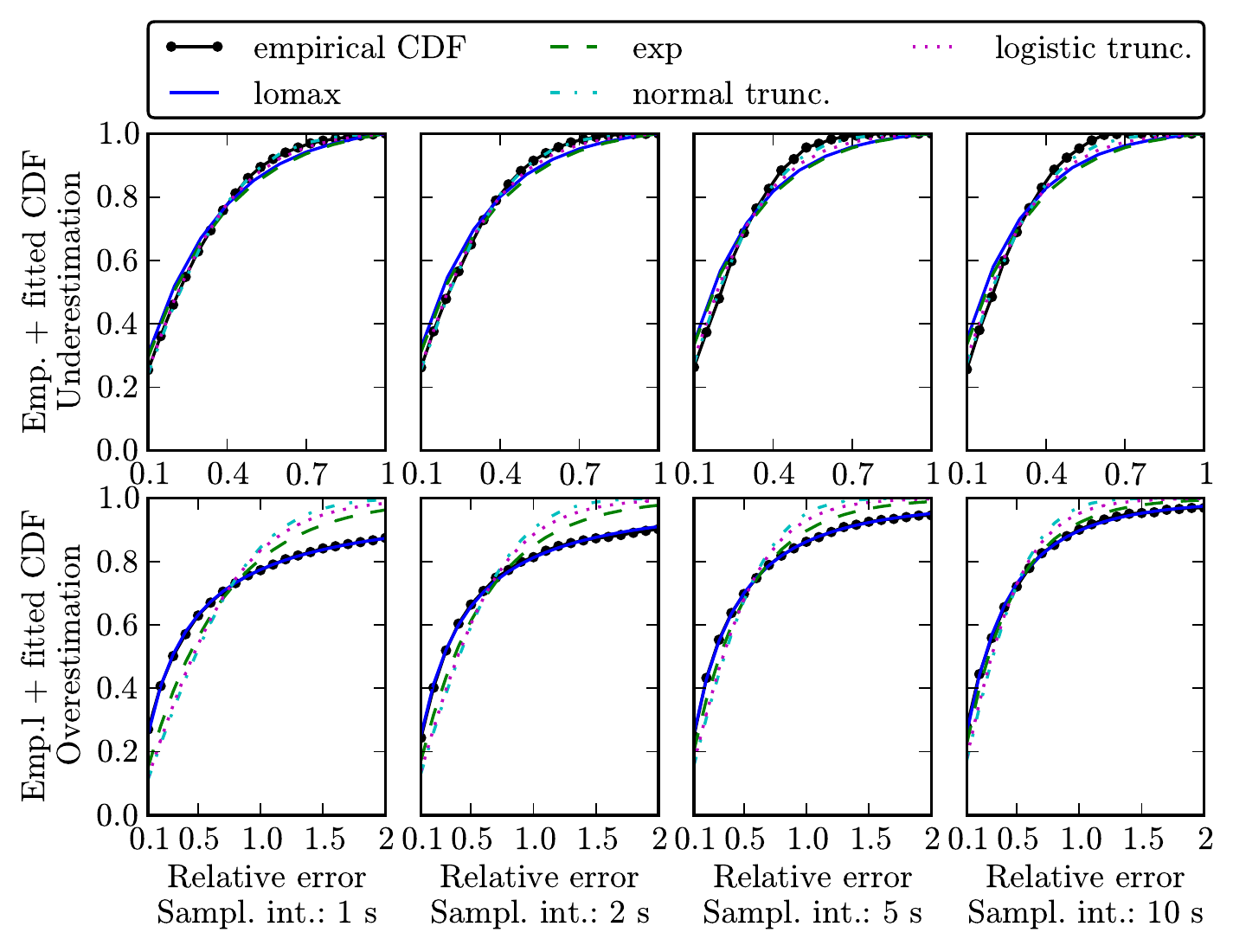}
\caption{Fitting distributions for relative prediction errors. See Section~\ref{sec:modeling} for details.}
\label{fig:distr_fitting_all}
\end{figure}

In order to evaluate the prediction accuracy, we use relative prediction error as metric, which we define as follows
\begin{equation}\label{eq:epsilon}
\epsilon\left(t_1,\,t_2\right)=\\
\frac{\lvert\max{\left(\hat{\rho}\left(t_1,\,t_2\right),\,\rho_{\min}\right)}-\max{\left(\rho\left(t_1,\,t_2\right),\,\rho_{\min}\right)}\rvert}{\max{\left(\rho\left(t_1,\,t_2\right),\,\rho_{\min}\right)}}\,,
\end{equation}
where $\hat{\rho}\left(t_1,\,t_2\right)$ is a throughput prediction for the time interval $\left[t_1,\,t_2\right]$. The maximum operator is necessary to avoid a distortion of results whenever $\rho\approx 0$ or $\hat{\rho}\leq 0$. We set $\rho_{\min}=10$ kbps.

We separately evaluate the overestimation and the underestimation error, due to the different error ranges of $(0,\infty)$ and $(0, 1]$, respectively, and due to their different impact on the streaming client. An overestimation increases the risk of missing a playback deadline, resulting in a playback interruption, which has the strongest impact on \ac{QoE}. An underestimation decreases the risk of interruptions but at the same time also reduces the media bit rate. In all studied traces, the relative frequency of underestimation and overestimation was close to 50\% (see Section~\ref{sec:ue_oe} for more details).

We made the, somewhat unexpected, observation that the increasing complexity of prediction methods such as \ac{ARIMA} or machine learning based approaches does not improve prediction accuracy. Our conclusion was that accounting for the trend in the data does not help decreasing the prediction error. A potential explanation might be the negative auto-correlation of throughput time series after differencing, depicted in Figure~\ref{fig:trace_statistics}. In particular, we observed that the median of the relative overestimation error of all studied methods is strictly greater than that of \ac{SMA}:1:ar. Therefore, in the following, we present results focusing on three simple methods: \ac{SMA}, linear extrapolation, and double exponential smoothing (Holt-Winters). 
%For Holt-Winters, the parameters are tuned for every prediction anew. 
%For the complete results, as well as a detailed description of methods used, please refer to the extended version of this paper. 

In the first step, we use for evaluation the complete set of measurements from all collected traces. The results are shown in Figures~\ref{fig:err_rel_cdf_all_traces_as_one_00}, \ref{fig:err_rel_cdf_all_traces_as_one_01}, and \ref{fig:err_rel_cdf_all_traces_as_one_02}.
Figure~\ref{fig:err_rel_cdf_all_traces_as_one_00} compares \ac{SMA} using different numbers of past measurements, computed from non-overlapping time intervals.
%Note that past measurements are computed from non-overlapping sampling intervals. That is, at time $t$, the most recent measurement contains statistics from the interval $[t-T_s, t]$, where $T_s$ is the duration of the sampling interval in seconds, while the second measurement contains statistics from $[t-2T_s, t-T_s]$, rather than $[t-T_s-1, t-1]$, etc. (Recall that the traces contain statistics computed for sliding intervals moving at 1 second steps.)
We observe that the overestimation error is smallest for \ac{SMA} with only one past measurement, while the underestimation error improves when using more measurements.
%Since the improvement from going from 5 to 10 past measurements is very small, we restrain from evaluating even higher values. 
Consequently, in the following, we use SMA:1 as reference for the overestimation error and SMA:10 for the underestimation error.
Figure~\ref{fig:err_rel_cdf_all_traces_as_one_01} compares linear extrapolation with the respective reference method. We observe that two past measurements provide worst results, improving with the increasing number of past measurements but always remaining below the performance of \ac{SMA}. Finally, Figure~\ref{fig:err_rel_cdf_all_traces_as_one_02} compares Holt-Winters with \ac{SMA}. Here, we observe the same situation as with linear extrapolation. It is also worth noting that Holt-Winters has a much higher computational complexity due to the optimization involved in tuning its configuration parameters for every new prediction.

%While Figure~\ref{fig:err_rel_cdf_all_traces_as_one} compares results for the joined set of collected measurements, we now turn our attention o the results for individual traces, depicted in Figure~\ref{fig:compare_quantiles}. 
Figure~\ref{fig:compare_quantiles} shows results for individual traces. In order to present them in a compact way,
%instead of showing for each trace the complete \ac{ECDF}, as in Figure~\ref{fig:err_rel_cdf_all_traces_as_one}, 
this figure shows for each trace only three selected points of the \ac{ECDF}, the fraction of measurements resulting in a relative error smaller than 0.2, 0.5, and 1.0, respectively. Each point on a subfigure corresponds to an individual trace so that the figures can be interpreted as \acp{ECDF} over individual traces (where smaller values indicate higher prediction accuracy).
The first two rows show results for a timescale of 2 seconds. The third and fourth rows for 5 seconds. The last two rows for 10 seconds. The first column shows for each trace the fraction of measurements with a relative error below 0.2, the second column below 0.5, and the last column below 1.0. The three missing subfigures are omitted since the relative error in the case of underestimation never exceeds $1.0$.
To give an example, each point in the subfigure in the top row first column shows the fraction of underestimations in a particular trace with a relative error of less than 0.2.
We observe that the per-trace comparison of prediction accuracy still indicates that in all cases, \ac{SMA} performs better than the more sophisticated methods.

From Figures~\ref{fig:err_rel_cdf_all_traces_as_one_00}, \ref{fig:err_rel_cdf_all_traces_as_one_01}, \ref{fig:err_rel_cdf_all_traces_as_one_02}, and~\ref{fig:compare_quantiles} we observe that with the reference method, a significant number of predictions result in a relatively small prediction error of below 20\%. For example, in some traces, even on the timescale of 10 seconds, approximately 80\% of predictions have an error smaller than 20\%, while almost 100\% of predictions have an error smaller than 50\%. A relative error of this magnitude could, in principle, be accounted for by a fixed safety margin, that is, by always selecting a media bit rate which is by 20\% smaller than the predicted throughput. There are, however, also "bad" traces, where more than 40\% of overestimated predictions have a relative error of greater than 50\%, while more than 20\% of predictions still have an error greater than 100\%, with even higher values for the timescale of 2 seconds. 

Setting a high fixed safety margin to account for "bad" traces would result in significant underutilization of network resources, lower media bit rate and thus lower \ac{QoE} in the "well-behaving" traces, while selecting a low fixed safety margin would increase the total re-buffering time in the "bad" traces, which, again, has a dramatic impact on \ac{QoE}. Therefore, in the following section, we focus our attention on approaches to modeling and estimating the prediction error.

\section{Error modeling}\label{sec:modeling}

\begin{figure}[t]
\centering
\includegraphics[scale=1]{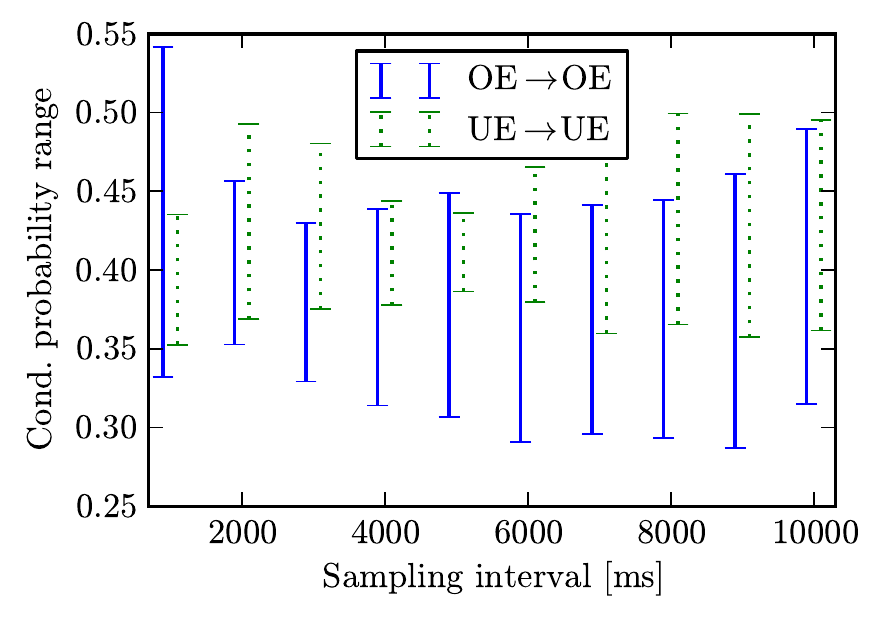}
\caption{Conditional probability ranges (over traces) for underestimations (overestimations), given that previous prediction was an underestimation (overestimation).}
\label{fig:ue_oe_prob}
\end{figure}

In order to be able to use throughput predictions for adaptation decisions in the most efficient way, we turn out attention to modeling and estimating the relative prediction error. The approach we follow is to determine, which type of distributions fits well the \ac{ECDF} of the prediction error. The developed adaptation algorithm for \ac{HALS}, presented in Chapter~\ref{sec:adaptation} leverages the obtained results and estimates the parameters of this distribution for individual streaming sessions, or repeatedly throughout a streaming session.

We use the following distributions: exponential, normal, logistic, and Lomax (shifted Pareto)~\cite{Johnson1994}. For the underestimation error, distributions are truncated to the range $[0, 1]$, for the overestimation error to the range $[0, \infty)$. The \ac{CDF} $F_{\text{tr}}(\cdot)$ of a distribution truncated to $[a,\,b]$ is obtained from the original \ac{CDF} $F(\cdot)$ as $$F_{\text{tr}}(x)=\frac{F(x)-F(a)}{F(b)-F(a)},\;\;x\in[a,\,b]\,.$$

We fit a distribution to the data by minimizing the squared distance ($L^2$-norm) between its \ac{CDF} and the truncated \ac{ECDF}. The \ac{ECDF} is truncated in order to make the fit more precise in the range which is relevant for adaptive streaming clients. The \ac{ECDF} of the overestimation error is truncated to the range $[0.1, 2.0]$, in the case of underestimation to the range $[0.1, 1.0]$. Afterwards, Kolmogorov-Smirnov test is used to verify the goodness of fit~\cite{Hollander2014}.

The results are shown in Figure~\ref{fig:distr_fitting_all}. The \acp{CDF} are fitted to \acp{ECDF} over the joined set of measurements from all traces. It turns out that the overestimation error is extremely well represented by a Lomax distribution. The underestimation error is best represented by a truncated normal distribution except for the sampling interval of 1 second, where the truncated logistic distribution has a slightly better Kolmogorov-Smirnov distance. These findings are consistent with those obtained by fitting the prediction errors from individual traces, which are omitted here.

% Note that the fact that the data is best matched by a Lomax distribution does not necessarily justifies the claim that they are power-low distributed.

% TODO: visualize, how this distributions vary over the different traces that we collected. (scatter plots of quantiles or something like that? use [0.2, 0.5, 1.0] quantiles).

% TODO: check if distribution fitting results change if we do or don't exclude outliers

\section{Underestimation and overestimation probabilities}\label{sec:ue_oe}

Since underestimation and overestimation errors have different ranges, $\left[0,\,1\right]$, and $\left[0,\,\infty\right)$, and due to their different impact on the operation of a streaming client, we separately study the probabilities for underestimation and overestimation occurrence. In the following, we limit our presentation to \ac{SMA}:1:ar.

In particular, we observed that in all studied traces the probability for occurrences of underestimations and overestimation are extremely well balanced on all timescales. Both occur in approximately $50\%\pm 2\%$ of predictions in a trace. However, it turns out that they exhibit significant temporal correlation. In particular, their conditional probabilities, if we take into account the nature of the last prediction error, are, in most cases, significantly below 50\%. In particular, the probability to encounter two underestimations or two overestimations in a row goes down to as low as 30\% for some traces on some timescales. This observation is directly related to the distinct negative correlation of the throughput process after differencing, depicted in Figure~\ref{fig:trace_statistics}. It, once again, highlights the strong variability of the throughput process on different timescales.

To provide an illustration, Figure~\ref{fig:ue_oe_prob} shows conditional probabilities for underestimation (overestimation), given that the previous prediction was an underestimation (overestimation) as well. The figure shows maximum and minimum values over all traces. We will make use of this result when computing rebuffering probabilities for video quality adaptation, presented in Section~\ref{sec:adaptation}.

\chapter{Prediction-based video adaptation}
\label{sec:adaptation}

% Note that predicted throughput values overestimate what the client really can use for download since a client sometimes have to delay subsequent segment downloads if a segment is not yet available.

%Selecting the quality for the next segment without considering the following ones might leave the client with a very low buffer level after finishing the download or might force the client to 

% Figure of the state diagram?

% Mention: assume segment sizes to be known or use prediction from Karim's paper.

% Even though a typical size of a video segment in \ac{HAS} constitutes one or few seconds, it is important to take into account not only the very next segment to be downloaded but also some of the subsequent segments. Otherwise, a client might decide to download a segment in a higher quality than can be sustained by the network, leaving little or no time to download subsequent segments before their playback deadlines.

In this chapter, we present our design of a novel prediction-based adaptation algorithm for live streaming that takes into account throughput predictions on different timescales, along with an estimation of prediction accuracy, and heuristically maximizes a \ac{QoE}-based utility function defined in the following. 
%The algorithm has a set of configuration parameters denoted by \kmi{$\left\{\alpha_{\text{mmbr}},\,\alpha_{\text{cdf}},\,\alpha_{\text{ur}}\right\}$} that will be defined in the following and that are summarized in \kmi{Table~\ref{tab:algo_config}}.

\section{General idea}

Our idea is, prior to each segment download, to compare potential future adaptation trajectories and to select the one that maximizes \ac{QoE}. In order to evaluate \ac{QoE} of an adaptation trajectory, we define a utility function depending on three factors that we call subutilities: (i) probability that a segment misses its playback deadline and thus the streaming session enters a rebuffering period, (ii) the distortion of video, evaluated by means of \ac{PSNR}, and (iii) the number and amplitude of representation changes. The utility is computed from individual subutilities based on results on their relative importance, available in the literature. In particular, rebuffering subutility appears as a multiplicative factor and plays the role of an upper bound on the total utility, reflecting its strong impact on \ac{QoE}. The other factor is a weighted sum of subutilities representing distortion and quality fluctuations that allows to resolve their trade-off in a configurable way. In order to compute the defined utility, in particular, the probability that a segment misses its playback deadline, we use the results of our study of \ac{TCP} throughput predictability from Chapter~\ref{sec:prediction}.

As defined in Chapter~\ref{sec:system_model}, the download of segment $i$ starts at $t_i^r$ and has to be completed until $t_i^p>t_i^r$. At $t_i^r$, the client either just finished downloading the previous segment $i-1$, or segment $i$ just became available at the server. Now, the client has to select a representation $j\in\{1,\ldots,|\mathcal{R}|\}$ for segment $i$. 

Parallel to downloading segments, every second, the client is recomputing throughput predictions for the following $1,2,\ldots,T_{\max}$ seconds. We denote by $t_i^\pi\leq t_i^r$ the most recent time when predictions were computed. That is, predictions are available for time intervals $\left[t_i^\pi,\,t_i^\pi+T\right]$, with $T\in\left\{1,2,\ldots,T_{\max}\right\}$. Assume that $l\geq i$ is the segment with the latest playback deadline that is still within the prediction horizon, that is, $l=\min\left\{l'\geq i \;|\; t_{l'}^p \leq t_i^\pi + T_{\max}\right\}$.

Let $\mathcal{T}_{il}=\left(s_i,\ldots,s_{l}\right)$ denote a vector of segment sizes, representing an adaptation trajectory, and $\mathfrak{T}_{il}$ the set of possible adaptation trajectories $\mathcal{T}_{il}$. We define the utility of a trajectory as a function of three subutilities, corresponding to estimated rebuffering probability, video quality, and video quality fluctuations, respectively:
\begin{equation*}
U(\mathcal{T}_{il}) = 
U_{\text{RB}}(\mathcal{T}_{il}) \cdot \left(\alpha_{\text{Q}}\cdot U_{\text{Q}}(\mathcal{T}_{il}) + \left(1-\alpha_{\text{Q}}\right)\cdot U_{\text{QF}}(\mathcal{T}_{il})\right)\,,
\end{equation*}
where $\alpha_{\text{Q}}\in\left[0,\,1\right]$ is a configuration parameter, determining the relative weighting of overall mean video signal quality versus quality fluctuations. The codomains of the subutilities, and thus of the utility, is the continuous interval $[0, 1]$. The multiplicative nature of rebuffering subutility reflects its strong impact on \ac{QoE}, especially for live streaming. According to a large-scale study of user engagement (time before the user quits a streaming session), it is always better to drop video quality than to let the streaming stall~\cite{Conviva2014}.

Let $\mathcal{T}^*_{il}\in\mathfrak{T}_{il}$ be the adaptation trajectory that maximizes $U(\cdot)$. Once identified, the client downloads segment $i$ from the representation it has in $\mathcal{T}^*_{il}$. Note that the representations for segments $i'>i$ might later be selected differently than in $\mathcal{T}^*_{il}$. The reason that the optimization still has to be performed over $\mathfrak{T}_{il}$ and not over $\mathfrak{T}_{ii}$ is that otherwise the state of the client after downloading segment $i$ would not be part of the optimization.
In this case, a client might, e.g., choose to change to a higher quality even though chances are high that he will have to switch back for subsequent segments. 
The computation of individual subutilities is explained in the following sections.

%\kmi{TODO: describe how we compute $\mathcal{T}^*$??}

\section{Rebuffering subutility}

The rebuffering subutility is a function of the probability that any segment in the given adaptation trajectory will miss its playback deadline. It attains 1 if the probability is 0, and decreases exponentially, at a configurable rate, until it attains 0 when the rebuffering probability reaches 1. In order to compute the rebuffering probability, we leverage \ac{TCP} throughput predictions over a time horizon of up to 10 seconds. Further, we use an estimation of probability that the next prediction will result in an underestimation or overestimation. Finally, we use an estimation of the \ac{CDF} of the relative underestimation and relative overestimation error from a configurable amount of past predictions.

We denote by $\hat{\rho}_{ik}=\hat{\rho}\left(t_i^\pi,\,t_i^\pi+T\right)$ the predicted throughput for the smallest interval $\left[t_i^\pi,\,t_i^\pi+T\right]$, $T\in\left\{1,2,\ldots,T_{\max}\right\}$, that contains $\left[t_i^r,\,t_k^p\right]$, for a $k\in\{i,\ldots,l\}$. The corresponding measured throughput shall be denoted by $\rho_{ik}$. $\epsilon_{ik}=\epsilon\left(t_i^\pi,\,t_i^\pi+T\right)$ shall denote the relative prediction error for $\hat{\rho}_{ik}$, as defined in~\eqref{eq:epsilon}. Further, we denote by $\Phi_{ik}^u\left(\epsilon_{ik}\right)$ and $\Phi_{ik}^o\left(\epsilon_{ik}\right)$ the estimated \ac{CDF} of the underestimation and overestimation errors for $\hat{\rho}_{ik}$, respectively, computed at $t_i^\pi$. The type of distribution shall be selected based on results from Chapter~\ref{sec:prediction}, while distribution parameters are estimated for each streaming session individually by minimizing the squared distance between the \ac{CDF} of the selected distribution and \ac{ECDF} of prediction errors collected over past $\alpha_{\text{cdf}}\cdot T$ seconds, as described in more details in Chapter~\ref{sec:prediction}. $P_{ik}^u\in[0, 1]$ shall denote the probability of underestimation, that is $P\left[\hat{\rho}_{ik}<\rho_{ik}\right]$, estimated over past $\alpha_{\text{cdf}}\cdot T$ seconds, as described in Section~\ref{sec:ue_oe}. $\alpha_{\text{cdf}}$ is a configuration parameters, determining the amount of past measurements used to estimate error distributions and underestimation/overestimation probabilities.

To simplify notation, similar to~\eqref{eq:epsilon}, we shall define
\begin{equation*}
\tilde{\epsilon}_{ik}=\frac{\max{\left(\hat{\rho}_{ik},\rho_{\min}\right)}-\max{(\rho_{ik},\rho_{\min})}}{\max{(\rho_{ik}, \rho_{\min})}}\,,
\end{equation*}
which is the relative prediction error without taking the absolute value, that is, $\tilde{\epsilon}_{ik}\in(-1,\infty)$. The \ac{CDF} for $\tilde{\epsilon}_{ik}$ is then given by
\begin{equation*}
\Phi_{ik}\left(\tilde{\epsilon}_{ik}\right)=
\begin{cases}
P_{ik}^u \cdot \Phi_{ik}^u\left(\lvert\tilde{\epsilon}_{ik}\rvert\right) &\text{for}\;\tilde{\epsilon}_{ik}<0\\
P_{ik}^u + \left(1-P_{ik}^u\right) \cdot \Phi_{ik}^o\left(\lvert\tilde{\epsilon}_{ik}\rvert\right) &\text{otherwise}\,.
\end{cases}
\end{equation*}

The probability that a segment $l'\in\{i,\ldots,l\}$ in $\mathcal{T}_{il}$ will miss its playback deadline can now be estimated by
\begin{equation*}
P\left[\frac{\sum_{i'=i}^{l'}{s_{i'}}}{t_{l'}^p-t_i^r}\leq\frac{\hat{\rho}_{il'}}{1+\epsilon_{il'}}\right]=\Phi_{il'}\left(\frac{\hat{\rho}_{il'}\left(t_{l'}^p-t_i^r\right)}{\sum_{i=1}^{l'}{s_i}}-1\right)\,.
\end{equation*}
Then, the probability that any segment in $\mathcal{T}_{il}$ will miss its playback deadline and thus a rebuffering will occur can be estimated by
\begin{equation*}
P_{\text{RB}}\left(\mathcal{T}_{il}\right)=1-\sum_{l'=i}^{l}\Phi_{il'}\left(\frac{\hat{\rho}_{il'}\left(t_{l'}^p-t_i^r\right)}{\sum_{i'=i}^{l'}{s_{i'}}}-1\right)\,.
\end{equation*}

We define the rebuffering subutility to be exponentially decreasing in the probability for a segment to miss its playback deadline: 
\begin{equation*}
U_{\text{RB}}\left(\mathcal{T}_{il}\right) = 
\frac{e^{\alpha_{\text{RB}} P_{\text{RB}}(\mathcal{T}_{il})}-e^{\alpha_{\text{RB}}}}{1-e^{\alpha_{\text{RB}}}}\,,
\end{equation*}
which is the exponential function shifted and rescaled to pass through $(0,1)$ and $(1,0)$. Configuration parameter $\alpha_{\text{RB}}<0$ can be used to tune the slope of the function. For $\alpha_{\text{RB}}\rightarrow 0$, the function converges to the linear function $f(x)=1-x$. For $\alpha_{\text{RB}}\rightarrow -\infty$, the function converges to 
\begin{equation*}
f(x)=\begin{cases}1 &\text{if}\;x=0\\0 &\text{otherwise}\,.\end{cases}
\end{equation*}

\section{Video quality subutility}\label{sec:mbr_subutility}

In order to quantify the video quality of a given adaptation trajectory, we evaluate its \ac{PSNR} and map it linearly to the interval $\left[0,\,1\right]$. Although \ac{PSNR} does not adequately represent \ac{QoE}, it can serve as an indicator of the distortion due to the compression applied to create a representation. The \ac{PSNR} of the representation with lowest quality is mapped to 0, while the \ac{PSNR} of the representation with highest quality is mapped to 1. 

Let $\left\{\gamma_1,\ldots,\gamma_m\right\}$ be the \ac{PSNR} values of the representations in $\mathcal{R}$. The video quality subutility of segment $s_{ij}$ from representation $j$ shall be defined as 
\begin{equation*}
U_{\text{Q}}\left(s_{ij}\right) = \frac{\gamma_j-\gamma_1}{\gamma_m-\gamma_1}\,.
\end{equation*}
The video quality subutility of an adaptation trajectory shall be defined as mean video quality subutility computed over representations of individual segments
\begin{equation*}
U_{\text{Q}}\left(\mathcal{T}_{il}\right) = \frac{1}{l-i+1}\sum_{l'=i}^{l}U_{\text{Q}}\left(s_{l'}\right)\,.
\end{equation*}

\section{Quality fluctuations subutility}

We define the quality fluctuations subutility as one minus mean change in video quality between subsequent segments of an adaptation trajectory. In order to compute the quality fluctuations subutility for a trajectory $\mathcal{T}_{il}=(s_i,\ldots,s_l)$, we construct an auxiliary trajectory $\tilde{\mathcal{T}}_{il}=(\tilde{s}_i,\ldots,\tilde{s}_l)$, where $\tilde{s}_{i'}$ is the segment size of segment $i'$ if it would have been selected from the same representation as segment $i'-1$ in trajectory $\mathcal{T}_{il}$. That is, segment $s_{i'-1}$ and segment $\tilde{s}_{i'}$ are always from the same representation. With this auxiliary trajectory, quality fluctuations subutility can be expressed as
\begin{equation*}
U_{\text{QF}}\left(\mathcal{T}_{il}\right) = 1-\frac{1}{l-i+1}\sum_{l'=i}^{l}{\lvert U_{\text{Q}}\left(s_{l'}\right)-U_{\text{Q}}\left(\tilde{s}_{l'}\right)\rvert}\,.
\end{equation*}

\section{Tuning into the stream}\label{sec:tune_in}

When the client is about to join a live stream, he has to decide with which segment to start the download, which quality to select for the first segment, and when to display it to the user. The decision influences \ac{QoE} in several ways. It impacts the initial delay, the maximum buffer level attainable during the streaming session (see Section~\ref{sec:system_model} for details), and the initial video quality. Moreover, the choice is restricted by the maximum delay constraint that we assume is defined by the service provider or by the client profile. 

In the presented approach, our solution is to maximize the attainable buffer level to increase robustness against throughput fluctuations that might lead to rebuffering. Since live streaming requires a relatively low maximum delay, its sensitivity to throughput fluctuations is particularly high. In addition, rebuffering was shown to have dramatic impact on \ac{QoE}. 

We maximize the attainable buffer level by presenting the first segment to the user when its maximum playback deadline is reached, in opposite to displaying it immediately after it is downloaded. This creates a moderate initial delay, which was shown to be preferred by viewers to rebuffering. In addition, we download the first segment in lowest quality, in order to avoid the situation where it misses its playback deadline and has to be skipped, unnecessarily increasing the initial delay. Due to the typically small duration of individual segments, we assume that it has negligible impact on \ac{QoE}.

%At the beginning of each streaming session, due to the maximum delay constraint, there is a trade-off between start-up delay, maximum attainable buffer level, initial video quality, and initial risk of a buffer underrun. 

Assume that $t$ is the time when a user tunes into a live stream. We select the first segment $i_0$ to be downloaded as the oldest available segment whose playback deadline is at least $\tau$ seconds into the future, at lowest quality:
\begin{equation}\label{eq:i0}
i_0=\min\left\{i\geq 0\;|\; (i+1)\tau\leq t \wedge t_i^p \geq t + \tau \right\}\,.
\end{equation}
The intuition behind that is that it is reasonable to assume that the available network resources should at least support the download of a segment in lowest quality in less time than the segment duration. 

Upon completing the download, the client waits until $t_{i_0}^p=i_0\tau+\Delta^p$, before presenting it to the user, in order to maximize the attainable buffer level, as described in Chapter~\ref{sec:system_model}. If the first segment can be downloaded before its playback deadline, as expected, the start-up delay will thus lie in the interval $\left[\tau,\,\Delta^p-\tau\right]$, which can be seen by transforming~\eqref{eq:i0}, using $t_i^p=i\tau+\Delta^p$.

\section{Missing playback deadlines}

Whenever a segment cannot be downloaded before its playback deadline, its download is canceled, and a tune-in procedure, described in Section~\ref{sec:tune_in} is initiated.

% as described in Section~\ref{sec:tune_in}.

%After re-buffering, we never skip segments. We assume that the user prefers It is better to play out old content than not to play anything at all. Skipping might then happen if we are re-buffering, a segment arrives too late (missing maximum playback deadline), is still played out to avoid longer re-buffering, meanwhile more segments arrive, some of which already have missed their maximum playback deadlines and will be skipped.

\chapter{Adaptive streaming client evaluation}
\label{sec:evaluation}

%\kmi{PSNR from \cite{Wiegand2003}}
%\kmi{plot buffer levels}

\begin{figure}
\centering
\includegraphics[scale=1]{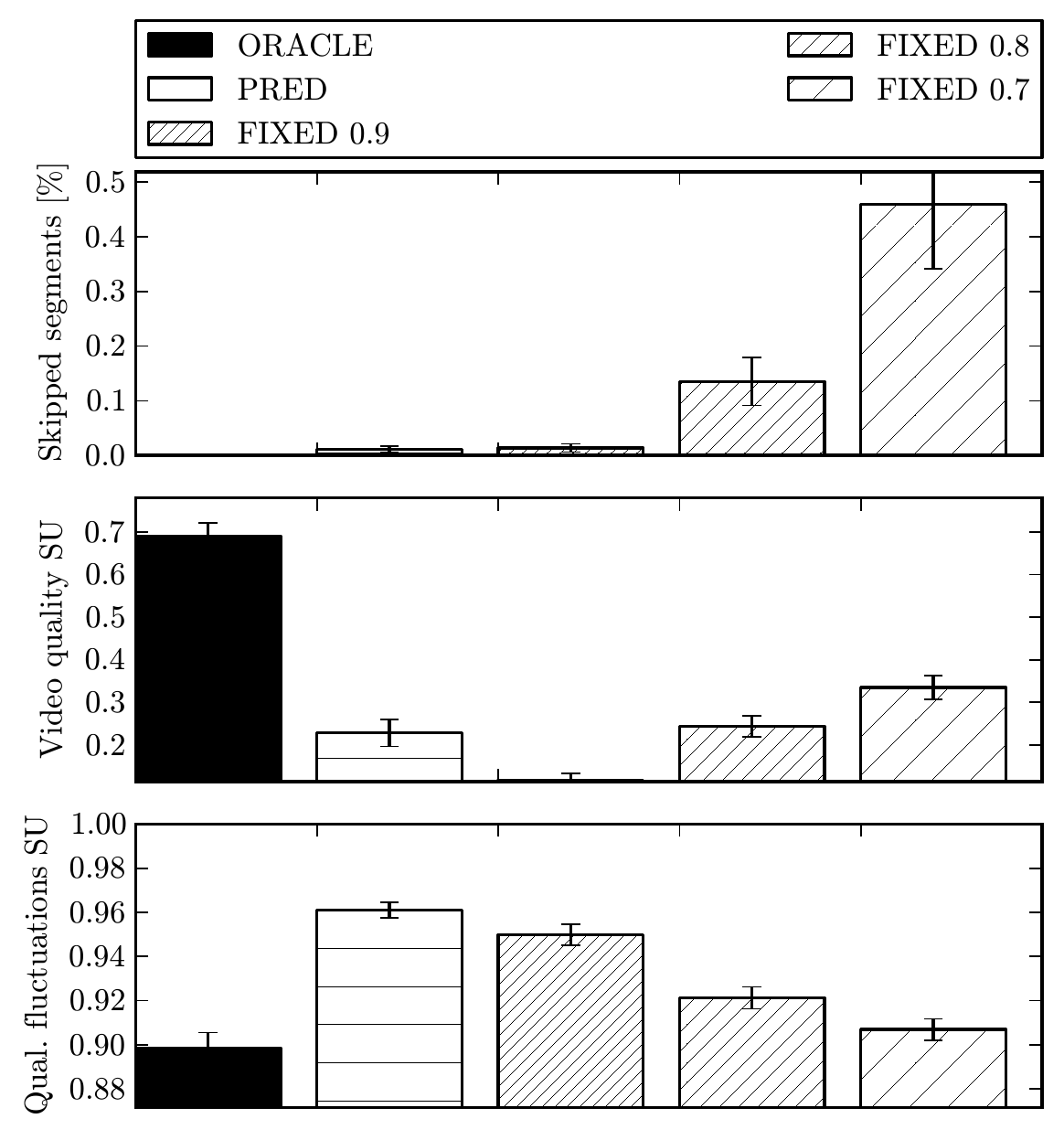}
\caption{Evaluation results: fraction of skipped segments (reduced by "unplayable" segments), video quality subutility, and quality fluctuations subutility. Vertical lines display confidence intervals for a confidence level of 0.9.}
\label{fig:algo_evaluation}
\end{figure}

In this section, we present the results of the evaluation of the proposed adaptation method based on simulations using throughput traces described in Section~\ref{sec:traces}. Its performance is compared to the baseline approach using a fixed margin between past average throughput and the \ac{MMBR} of the selected representation. It is further compared to the performance of an approach that has perfect knowledge of future throughput for a limited time horizon.

We use the following performance metrics: fraction of skipped segments, mean video quality subutility, and mean quality fluctuations subutility, computed over individual streaming sessions, lasting for the duration of individual traces. The videos for the evaluation are taken from the dataset provided by Lederer et al.~\cite{Lederer2012}. The segment duration equals 2 seconds. For each video, we are using 10 representations, with \acp{MMBR} ranging from 100 to 4200 kbps.

Some of the traces contain short periods when the throughput falls below the lowest available segment \ac{MMBR}, so that a certain fraction of segments inevitably has to be skipped, independent of the deployed adaptation strategy. In order to compute this fraction, we simulate an adaptation algorithm that always selects the lowest representation, leading to a video quality subutility of 0 and a quality fluctuations subutility of 1. In the following, when we present the fraction of skipped segments for a simulation run, we always subtract the fraction of such "unplayable" segments.

The performance of the developed approach is compared to a baseline approach that uses a fixed margin between past average throughput and the \ac{MMBR} of the selected representation. The margin is varied between 0.7 and 0.9. Effectively, this means that the baseline approach uses SMA:1:ar to predict throughput, assumes that the prediction is always an overestimate, and that the relative error is fixed.

We also compare to an oracle-based approach which has full knowledge about future throughput for a horizon of 10 seconds, and which selects for the next segment the highest representation that does not result in rebuffering within this time horizon.

We set the maximum playback delay to $\Delta^p_{\text{max}}=5$ seconds, which corresponds to a maximum transmission delay of $\Delta^p_{\text{max}}-\tau=3$ seconds, that is, 1.5 times the segment duration. Theoretically it is possible to further reduce the maximum transmission delay to equal segment duration, which, however, we consider infeasible since it would dramatically increase system's sensitivity to throughput fluctuations.

We set $\alpha_{\text{cdf}}=60$. That is, the error for a prediction 2 seconds into the future is computed by estimating \ac{CDF} parameters from past $2\alpha_{\text{cdf}}=120$ seconds. We set $\alpha_{\text{RB}}=-200$, which results in a steeply decreasing rebuffering subutility, since we define it our highest priority to avoid rebuffering, potentially sacrificing video quality and taking into account quality fluctuations. 

Finally, we set $\alpha_{\text{Q}}=0.6$, giving quality subutility a slightly higher weight for the overall utility, than the quality fluctuations subutility.

The results of the evaluation are depicted in Figure~\ref{fig:algo_evaluation}. With the proposed algorithm, the mean fraction of skipped segments, without "unplayable" segments, is measured at approximately $10^{-4}$. This means that on average, a segment that could be played if perfect knowledge about future throughput would be available is skipped every half an hour. Recall that our highest priority is to keep the rebuffering rate as low as possible. The study in~\cite{Conviva2014}, e.g., suggests that live stream viewers can be extremely sensitive to rebuffering and might quit a streaming session after experiencing a single rebuffering event after a long time of uninterrupted viewing. 

The mean quality subutility amounts to approximately one third of what is achieved given perfect knowledge about future throughput. The remaining two thirds constitute the inevitable tribute to the uncertainty of future throughput dynamics. Considering quality fluctuations, the proposed algorithm even achieves a higher subutility than the oracle approach.

We observe that in order to achieve a comparably small fraction of skipped segments using the fixed margin approach, a margin strictly greater than 0.9 is required (for exactly 0.9 the value is still slightly higher). At the same time, however, such a high margin results in significantly lower mean video quality and higher quality fluctuations. Further decreasing the margin downgrades the performance even more, leading to a fraction of skipped segments that is one order of magnitude higher.

Finally, we remark that in the present study we focused on network conditions with extremely high throughput fluctuations. Using the fixed margin approach on a link with moderate or low throughput fluctuations would further decrease resource utilization and thus further downgrade the performance. In the extreme case, if the throughput is constant, a fixed margin $x$ leads to a resource utilization of $1-x$. In contrast, since our algorithm dynamically estimates the relative prediction error, with constant throughput it is able to achieve a utilization close to 1. Thus, one of the strengths of the proposed method lies in its ability to perform equally well in very different network environments.

\chapter{Conclusion}
\label{sec:conclusion}

In the presented study, we proposed an adaptation algorithm for \ac{HALS}. It is based on the idea to compare potential future adaptation trajectories and to select the one maximizing \ac{QoE}. \ac{QoE} is evaluated based on a utility function depending on (i) the probability that a segment misses its playback deadline, (ii) the distortion of the video, and (iii) the number and amplitude of representation changes. 

In order to compute the defined utility, in particular, the probability that a segment misses its playback deadline, we studied predictability of \ac{TCP} throughput in wireless networks over timescales from 1 to 10 seconds. We evaluated different time series prediction methods using varying numbers of past throughput measurements. We demonstrated that the most na\"{\i}ve method, \ac{SMA}, outperforms more sophisticated methods on all timescales, independent of the specific throughput dynamics. We further observed that prediction accuracy strongly varies across studied traces. Consequently, we studied approaches to model the prediction error and to estimate it for individual streaming sessions. 
We demonstrated that the overestimation error is extremely well represented by the Lomax distribution~\cite{Johnson1994} on all considered timescales. The underestimation error is best represented by a truncated normal distribution except for the timescale of 1 second, where the truncated logistic distribution results in a slightly better Kolmogorov-Smirnov distance between the empirical and the theoretical \ac{CDF}. In addition, we found out that although underestimations and overestimations are balanced over the total duration of individual traces, they exhibit a strong temporal correlation that we used to further improve prediction accuracy.

Using obtained insights, the proposed adaptation algorithm takes into account throughput predictions and an estimation of the relative prediction error, in order to maximize the defined \ac{QoE}-based utility function. We evaluated the developed algorithm using collected throughput traces and showed that it outperforms the baseline approach which uses a fixed margin between past throughput and selected media bit rate.

Our ongoing and future work includes extending our collection of traces and including traces from mobile networks. It also includes studying the influence of ON/OFF patterns, generated by inter-request delays, on throughput prediction accuracy. Moreover, we are investigating how the prediction can be further improved by taking into account cross-layer information from \ac{TCP} and \ac{MAC} layers.
%Finally, we study how a client might dynamically tune the playback delay based on statistics of throughput in general and link outages in particular.

%Ongoing work includes extending the set of traces, as well as including traces from 3G/4G networks.

%In a live-streaming scenario, the client cannot continuously download data at maximum rate offered by \ac{TCP}. If the average download rate of preceding segments was higher than their average media bit rate, he might have to wait for the next segment to become available. In this case, the resulting \ac{TCP} flow follows a certain ON/OFF pattern. It is an ongoing work to investigate the influence of this pattern on prediction accuracy.

%Better prediction methods

%Calculate optimal trajectories for the live-streaming scenario, similar to our previous paper, to compare with.

%Improve underestimation/overestimation prediction.

%Single-step versus multi-step prediction.

% compute \ac{CDF} for the paper with gaussian predictions, std 10\%, for comparison.

% relate to the markov decision process paper computing offline a model based on distribution and show that our results how that this model based on overall distribution must perform poorly.

\acrodef{ADS}{Adobe Dynamic Streaming}
\acrodef{AIC}{Akaike Information Criterion}
\acrodef{AP}{Access Point}
\acrodef{ARIMA}{Autoregressive Integrated Moving Average}
\acrodef{BOWL}{Berlin Open Wireless Lab}
%\acrodef{BRAS}{Block-Request Adaptive Streaming}
\acrodef{CDF}{Cumulative Distribution Function} \acrodefplural{CDF}[CDF's]{Cumulative Distribution Functions}
\acrodef{CDN}{Content Delivery Network} \acrodefplural{CDN}[CDN's]{Content Delivery Networks}
\acrodef{CIF}{Context Influence Factor} \acrodefplural{CIF}[CIF's]{Context Influence Factors}
\acrodef{CSS}{Cubic Smoothing Splines}
\acrodef{CV}{Coefficient of Variation}
\acrodef{DASH}{Dynamic Adaptive Streaming over HTTP}
\acrodef{DSL}{Digital Subscriber Line}
\acrodef{ECDF}{Empirical Cumulative Distribution Function} \acrodefplural{ECDF}[ECDF's]{Empirical Cumulative Distribution Functions}
\acrodef{MPEG-DASH}{Dynamic Adaptive Streaming over HTTP}
\acrodef{GOP}{Group of Pictures}
\acrodef{HAS}{HTTP-Based Adaptive Streaming}
\acrodef{HALS}{HTTP-Based Adaptive Live Streaming}
\acrodef{HD}{High-Definition}
\acrodef{HLS}{Apple HTTP Live Streaming}
\acrodef{HTML}{Hypertext Markup Language}
\acrodef{HTTP}{Hypertext Transfer Protocol}
\acrodef{HIF}{Human Influence Factor} \acrodefplural{HIF}[HIF's]{Human Influence Factors}
\acrodef{IP}{Internet Protocol}
\acrodef{ITU}{International Telecommunication Union}
\acrodef{IVP}{Initial Value Problem}
\acrodef{LAN}{Local Area Network}
\acrodef{LMI}{Linear Matrix Inequality}
\acrodef{LOESS}{Locally Weighted Scatterplot Smoothing}
\acrodef{LOS}{Line-of-Sight}
\acrodef{LQR}{Linear Quadratic Regulator}
\acrodef{LTE}{Long-Term Evolution}
\acrodef{MAC}{Media Access Control}
\acrodef{MANET}{Mobile Ad-Hoc Network}
%\acrodef{MBRF}{Media Bit Rate Fluctuation} \acrodefplural{MBRF}[MBRF's]{Media Bit Rate Fluctuations}
\acrodef{MCNKP}{Multiple-Choice Nested Knapsack Problem}
\acrodef{MIMO}{Multiple Input Multiple Output}
\acrodef{MLE}{Maximum Likelihood Estimation}
\acrodef{MMBR}{Mean Media Bit Rate} \acrodefplural{MMBR}[MMBR's]{Mean Media Bit Rates}
\acrodef{MPC}{Model Predictive Control}
\acrodef{MPD}{Media Presentation Description}
\acrodef{MSE}{Mean Squared Error}
\acrodef{MSS}{Microsoft SmoothStreaming}
\acrodef{NAT}{Network Address Translation}
\acrodef{NCS}{Networked Control System}
\acrodef{NLOS}{Non-Line-of-Sight}
\acrodef{NUM}{Network Utility Maximization}
\acrodef{ODE}{Ordinary Differential Equation}
\acrodef{OFDM}{Orthogonal Frequency-Division Multiplexing}
\acrodef{PI}{Proportional-Integral}
\acrodef{PID}{Proportional-Integral-Derivative}
\acrodef{PSNR}{Peak Signal-to-Noise Ratio}
\acrodef{RTT}{Round-Trip Time}
\acrodef{TCP}{Transmission Control Protocol}
\acrodef{QoE}{Quality of Experience}
\acrodef{QoS}{Quality of Service}
\acrodef{SES}{Simple Exponential Smoothing}
\acrodef{SIF}{System Influence Factor} \acrodefplural{SIF}[SIF's]{System Influence Factors}
\acrodef{SINR}{Signal-to-Interference-plus-Noise Ratio}
\acrodef{SISO}{Single Input Single Output}
\acrodef{SMA}{Simple Moving Average}
\acrodef{SMC}{Sliding Mode Control}
\acrodef{SNS}{Social Networking Service} \acrodefplural{SNS}[SNS's]{Social Networking Services}
\acrodef{SSI}{Signal Strength Indicator}
\acrodef{UGC}{User-Generated Content}
\acrodef{URL}{Uniform Resource Locator} \acrodefplural{URL}[URL's]{Uniform Resource Locators}
\acrodef{VBR}{Variable Bit Rate}
\acrodef{VoD}{Video on Demand}
\acrodef{VSC}{Variable Structure Control}
\acrodef{WLAN}{Wireless Local Area Network} \acrodefplural{WLAN}[WLAN's]{Wireless Local Area Networks}
\acrodef{XAP}{Silverlight Application Package}

%\bibliographystyle{plain}
%\bibliography{library}
%\bibliographystyle{IEEEtran}
\bibliography{IEEEabrv,library}

\begin{thebibliography}{10}

\bibitem{radiotap}
{Radiotap}.
\newblock http://www.radiotap.org.

\bibitem{ITU-T_E.800}
{Definition of Terms Related to Quality of Service (ITU-T E.800)}.
\newblock {\em International Telecommunication Union Recommendation}, 2008.

\bibitem{ITUT_P10G100_Am2}
{Vocabulary for Performance and Quality of Service, Amendment 2: New
  Definitions for Inclusion in Recommendation ITU-T P.10/G.100}.
\newblock {\em International Telecommunication Union Recommendation}, 2008.

\bibitem{DASH2012}
{MPEG-DASH (ISO/IEC 23009-1)}.
\newblock {\em Standard}, 2012.

\bibitem{CiscoVNI2014}
{Cisco Visual Networking Index: Forecast and Methodology, 2013 - 2018}.
\newblock {\em Cisco, Inc. Report}, 2014.

\bibitem{comScore2014a}
{U.S. Digital Future in Focus}.
\newblock {\em comScore, Inc. Whitepaper}, 2014.

\bibitem{Conviva2014}
{Viewer Experience Report}.
\newblock {\em Conviva Report}, 2014.

\bibitem{Conviva2015}
{Internet TV: Bringing Control to Chaos}.
\newblock {\em Conviva Whitepaper}, 2015.

\bibitem{YouTubeStats_medium}
{YouTube Statistics}.
\newblock http://www.youtube.com/yt/press/statistics.html, 2015.

\bibitem{Aurrecoechea1998}
Cristina Aurrecoechea, Andrew~T. Campbell, and Linda Hauw.
\newblock {A Survey of QoS Architectures}.
\newblock {\em Multimedia Systems}, 6(3):138--151, 1998.

\bibitem{Balachandran2012}
Athula Balachandran, Vyas Sekar, Aditya Akella, Srinivasan Seshan, Ion Stoica,
  and Hui Zhang.
\newblock {A Quest for an Internet Video Quality-of-Experience Metric}.
\newblock In {\em In Proc. of ACM Workshop on Hot Topics in Networks
  (HotNets)}, Redmond, WA, USA, 2012.

\bibitem{Balachandran2013}
Athula Balachandran, Vyas Sekar, Aditya Akella, Srinivasan Seshan, Ion Stoica,
  and Hui Zhang.
\newblock {Developing a Predictive Model of Quality of Experience for Internet
  Video}.
\newblock In {\em Proc. of ACM SIGCOMM}, Hong Kong, 2013.

\bibitem{Broxton2011}
Tom Broxton, Yannet Interian, Jon Vaver, and Mirjam Wattenhofer.
\newblock {Catching a Viral Video}.
\newblock {\em Journal of Intelligent Information Systems}, 40(2):241--259,
  December 2013.

\bibitem{Carapinha2010}
Jorge Carapinha, Roland Bless, Christoph Werle, Konstantin Miller, Virgil
  Dobrota, Andrei~Bogdan Rus, Heidrun Grob-Lipski, and Horst Roessler.
\newblock {Quality of Service in the Future Internet}.
\newblock In {\em In Proc. of ITU-T Kaleidoscope}, Pune, India, 2010.

\bibitem{Hao2014}
Jia Hao, Roger Zimmermann, and Haiyang Ma.
\newblock {GTube: Geo-Predictive Video Streaming over HTTP in Mobile
  Environments}.
\newblock In {\em In Proc. of ACM Multimedia Systems Conference (MMSys)},
  Singapore, 2014.

\bibitem{Hollander2014}
Myles Hollander, Douglas~A. Wolfe, and Eric Chicken.
\newblock {\em {Nonparametric Statistical Methods}}.
\newblock Wiley, 2014.

\bibitem{Hossfeld2012a}
T.~Hossfeld, Sebastian Egger, Raimund Schatz, Markus Fiedler, Kathrin Masuch,
  and C.~Lorentzen.
\newblock {Initial Delay vs. Interruptions: Between the Devil and the Deep Blue
  Sea}.
\newblock In {\em In Proc. of Workshop on Quality of Multimedia Experience
  (QoMEX)}, Yarra Valley, Australia, 2012.

\bibitem{Hyndman2007}
Rob~J. Hyndman and Yeasmin Khandakar.
\newblock {Automatic Time Series Forecasting: The Forecast Package for R}.
\newblock {\em Journal of Statistical Software, University of California, Los
  Angeles, Department of Statistics}, 27(3):1--22, 2008.

\bibitem{Hyndman2005a}
Rob~J. Hyndman, Maxwell~L. King, Ivet Pitrun, and Baki Billah.
\newblock {Local Linear Forecasts Using Cubic Smoothing Splines}.
\newblock {\em Australian \& New Zealand Journal of Statistics}, 47(1):87--99,
  2005.

\bibitem{Jarnikov2011a}
Dmitri Jarnikov and Tanır \"{O}z\c{c}elebi.
\newblock {Client Intelligence for Adaptive Streaming Solutions}.
\newblock {\em Signal Processing: Image Communication}, 26(7):378--389, August
  2011.

\bibitem{Jiang2012}
Junchen Jiang, Vyas Sekar, and Hui Zhang.
\newblock {Improving Fairness, Efficiency, and Stability in HTTP-Based Adaptive
  Video Streaming with FESTIVE}.
\newblock In {\em In Proc. of ACM Conference on emerging Networking EXperiments
  and Technologies (CoNEXT)}, Nice, France, 2012.

\bibitem{Johnson1994}
Norman~Lloyd Johnson, Samuel Kotz, and Narayanaswamy Balakrishnan.
\newblock {\em {Continuous Univariate Distributions}}.
\newblock Wiley, 1994.

\bibitem{Krishnappa2013}
Dilip~Kumar Krishnappa, Divyashri Bhat, and Michael Zink.
\newblock {DASHing YouTube: An Analysis of Using DASH in YouTube Video
  Service}.
\newblock In {\em In Proc. of IEEE Conference on Local Computer Networks
  (LCN)}, Sydney, Australia, 2013.

\bibitem{Lederer2012}
Stefan Lederer, Christopher M\"{u}ller, and Christian Timmerer.
\newblock {Dynamic Adaptive Streaming over HTTP Dataset}.
\newblock In {\em In Proc. of ACM Multimedia Systems Conference (MMSys)},
  Chapel Hill, NC, USA, 2012.

\bibitem{Lewcio2011}
Blazej Lewcio, Benjamin Belmudez, Theresa Enghardt, and Sebastian M\"{o}ller.
\newblock {On the Way to High-Quality Video Calls in Future Mobile Networks}.
\newblock In {\em In Proc. of International Workshop on Quality of Multimedia
  Experience (QoMEX)}, Mechelen, Belgium, 2011.

\bibitem{Li2013a}
Baochun Li, Zhi Wang, Jiangchuan Liu, and Wenwu Zhu.
\newblock {Two Decades of Internet Video Streaming: A Retrospective View}.
\newblock {\em ACM Transactions on Multimedia Computing, Communications, and
  Applications}, 9(1s):1--20, 2013.

\bibitem{Liu2011}
Chenghao Liu, Imed Bouazizi, and Moncef Gabbouj.
\newblock {Rate Adaptation for Adaptive HTTP Streaming}.
\newblock In {\em In Proc. of ACM Multimedia Systems Conference (MMSys)}, San
  Jose, CA, USA, 2011.

\bibitem{Liu2012a}
Xi~Liu, Florin Dobrian, Henry Milner, Junchen Jiang, Vyas Sekar, Ion Stoica,
  and Hui Zhang.
\newblock {A Case for a Coordinated Internet Video Control Plane}.
\newblock In {\em In Proc. of ACM SIGCOMM}, Helsinki, Finland, 2012.

\bibitem{Liu2014a}
Yan Liu and Jack Y.~B. Lee.
\newblock {On Adaptive Video Streaming with Predictable Streaming Performance}.
\newblock In {\em In Proc. of IEEE Global Communications Conference
  (GLOBECOM)}, Austin, TX, USA, 2014.

\bibitem{MillerK2012}
Konstantin Miller, Emanuele Quacchio, Gianluca Gennari, and Adam Wolisz.
\newblock {Adaptation Algorithm for Adaptive Streaming over HTTP}.
\newblock In {\em Proc. of the Packet Video Workshop}, Munich, Germany, 2012.

\bibitem{Mok2012}
Ricky K.~P. Mok, Xiapu Luo, Edmond W.~W. Chan, and Rocky K.~C. Chang.
\newblock {QDASH: A QoE-Aware DASH System}.
\newblock In {\em In Proc. of ACM Multimedia Systems Conference (MMSyS)},
  Chapel Hill, NC, USA, 2012.

\bibitem{O'Reilly2007}
Tim O'Reilly.
\newblock {What Is Web 2.0: Design Patterns and Business Models for the Next
  Generation of Software}.
\newblock {\em Communications \& Strategies}, 1:17--37, 2007.

\bibitem{Pessemier2013}
Toon~De Pessemier, Katrien~De Moor, Wout Joseph, Lieven~De Marez, and Luc
  Martens.
\newblock {Quantifying the Influence of Rebuffering Interruptions on the
  User’s Quality of Experience During Mobile Video Watching}.
\newblock {\em IEEE Transactions on Broadcasting}, 59(1):47--61, 2013.

\bibitem{Quan2008}
Huynh-Thu Quan and Mohammed Ghanbari.
\newblock {Temporal Aspect of Perceived Quality in Mobile Video Broadcasting}.
\newblock {\em IEEE Transactions on Broadcasting}, 54(3):641--651, 2008.

\bibitem{Reiter2014}
Ulrich Reiter, Kjell Brunnstr\"{o}m, Katrien~De Moor, Mohamed-Chaker Larabi,
  Manuela Pereira, Antonio Pinheiro, Junyong You, and Andrej Zgank.
\newblock {Factors Influencing Quality of Experience}.
\newblock In {\em Quality of Experience}, pages 55--74. Springer International
  Publishing, 2014.

\bibitem{Riiser2012}
Haakon Riiser, Tore Endestad, Paul Vigmostad, Carsten Griwodz, and Pal
  Halvorsen.
\newblock {Video Streaming Using a Location-Based Bandwidth-Lookup Service for
  Bitrate Planning}.
\newblock {\em ACM Transactions on Multimedia Computing, Communications, and
  Applications}, 8(3):1--19, 2012.

\bibitem{Seufert2014}
Michael Seufert, Sebastian Egger, Martin Slanina, Thomas Zinner, Tobias
  Hossfeld, and Phuoc Tran-Gia.
\newblock {A Survey on Quality of Experience of HTTP Adaptive Streaming}.
\newblock {\em IEEE Communications Surveys \& Tutorials, to appear}, 2014.

\bibitem{Singh2012a}
Kamal~Deep Singh, Yassine Hadjadj-Aoul, and Gerardo Rubino.
\newblock {Quality of Experience Estimation for Adaptive HTTP/TCP Video
  Streaming Using H.264/AVC}.
\newblock In {\em In Proc. of IEEE Consumer Communications and Networking
  Conference (CCNC)}, Las Vegas, NV, USA, 2012.

\bibitem{Sodagar2011}
Iraj Sodagar.
\newblock {The MPEG-DASH Standard for Multimedia Streaming Over the Internet}.
\newblock {\em IEEE Multimedia}, 18(4):62--67, 2011.

\bibitem{Song2014}
Wei Song and Dian~W. Tjondronegoro.
\newblock {Acceptability-Based QoE Models for Mobile Video}.
\newblock {\em IEEE Transactions on Multimedia}, 16(3):738--750, 2014.

\bibitem{Stockhammer2011a}
Thomas Stockhammer.
\newblock {Dynamic Adaptive Streaming over HTTP -- Standards and Design
  Principles}.
\newblock In {\em In Proc. of ACM Multimedia Systems Conference (MMSys)}, San
  Jose, CA, USA, 2011.

\bibitem{Gigaom2014}
Paul Sweeting.
\newblock {Video in 2014: Going Live and Over the Top}.
\newblock {\em Gigaom Research Report}, 2014.

\bibitem{Tian2012}
Guibin Tian and Yong Liu.
\newblock {Towards Agile and Smooth Video Adaptation in Dynamic HTTP
  Streaming}.
\newblock In {\em In Proc. of ACM Conference on emerging Networking EXperiments
  and Technologies (CoNEXT)}, Nice, France, 2012.

\bibitem{Wang2008}
Bing Wang, Jim Kurose, Prashant Shenoy, and Don Towsley.
\newblock {Multimedia streaming via TCP: An analytic performance study}.
\newblock {\em ACM Transactions on Multimedia Computing, Communications, and
  Applications}, 4(2):1--22, May 2008.

\bibitem{Yin2014}
Xiaoqi Yin, Vyas Sekar, and Bruno Sinopoli.
\newblock {Toward a Principled Framework to Design Dynamic Adaptive Streaming
  Algorithms over HTTP}.
\newblock In {\em In Proc. of ACM Workshop on Hot Topics in Networks
  (HotNets)}, Los Angeles, CA, USA, 2014.

\bibitem{Yitong2013}
Liu Yitong, Shen Yun, Mao Yinian, Liu Jing, Lin Qi, and {Yang Dacheng}.
\newblock {A Study on Quality of Experience for Adaptive Streaming Service}.
\newblock In {\em In Proc. of IEEE International Conference on Communications
  (ICC) Workshops}, 2013.

\bibitem{Zhang2001}
Qian Zhang, Guijin Wang, Wenwu Zhu, and Ya-Qin Zhang.
\newblock {Robust Scalable Video Streaming over Internet with Network-Adaptive
  Congestion Control and Unequal Loss Protection}.
\newblock In {\em In Proc. of Packet Video Workshop}, Kyongju, Korea, 2001.

\end{thebibliography}

%\bibliography{biblio}

\end{document}